\documentclass[lettersize,journal, onecolumn]{IEEEtran}
\usepackage{amsmath,amsfonts}
\usepackage{algorithmic}
\usepackage{algorithm}
\usepackage{array}
\usepackage{textcomp}
\usepackage{stfloats}
\usepackage{url}
\usepackage{verbatim}
\usepackage{graphicx}
\usepackage{cite}
\usepackage{amssymb}
\usepackage{mathtools}
\usepackage{multicol}
\usepackage{multirow,tabularx}
\usepackage{booktabs}
\usepackage{caption}
\usepackage{subfigure}
\usepackage{hyperref}
\usepackage{algorithm}
\usepackage{algorithmic}

\usepackage{algorithm}
\usepackage{algorithmic}

\usepackage{amsmath}
\usepackage{amssymb}
\usepackage{amsthm}
\usepackage{multirow,tabularx}
\usepackage{multicol}
\usepackage{booktabs}
\usepackage{enumitem}
\usepackage[dvipsnames]{xcolor}
\usepackage[para,online,flushleft]{threeparttable}

\usepackage{amsmath}
\usepackage{amssymb}
\usepackage{amsthm}
\usepackage{multirow,tabularx}
\usepackage{multicol}
\usepackage{booktabs}
\usepackage[dvipsnames]{xcolor}
\usepackage[dvipsnames]{xcolor}
\begin{document}

\title{Fortifying Fully Convolutional Generative Adversarial Networks for Image Super-Resolution Using Divergence Measures}

\author{Arkaprabha Basu, Kushal Bose, Sankha Subhra Mullick, Anish Chakrabarty, and Swagatam Das \IEEEcompsocitemizethanks{\IEEEcompsocthanksitem Arkaprabha Basu, Kushal Bose, Sankha Subhra Mullick, Anish Chakrabarty and Swagatam Das (swagatam.das@isical.ac.in) are with the Electronics and Communication Sciences Unit (ECSU), Dolby Laboratories, India and Statistics and Mathematics Unit (SMU), Indian Statistical Institute, Kolkata, India \protect
\IEEEcompsocthanksitem Corresponding author: Swagatam Das.}}



\maketitle

\begin{abstract}
Super-Resolution (SR) is a time-hallowed image processing problem that aims to improve the quality of a Low-Resolution (LR) sample up to the standard of its High-Resolution (HR) counterpart. We aim to address this by introducing Super-Resolution Generator (SuRGe), a fully-convolutional Generative Adversarial Network (GAN)-based architecture for SR. We show that distinct convolutional features obtained at increasing depths of a GAN generator can be optimally combined by a set of learnable convex weights to improve the quality of generated SR samples. In the process, we employ the Jensen–Shannon and the Gromov-Wasserstein losses respectively between the SR-HR and LR-SR pairs of distributions to further aid the generator of SuRGe to better exploit the available information in an attempt to improve SR. Moreover, we train the discriminator of SuRGe with the Wasserstein loss with gradient penalty, to primarily prevent mode collapse. The proposed SuRGe, as an end-to-end GAN workflow tailor-made for super-resolution, offers improved performance while maintaining low inference time. The efficacy of SuRGe is substantiated by its superior performance compared to 28 state-of-the-art contenders on 10 benchmark datasets.
\end{abstract}


\section{Introduction}
A Low-resolution (LR) image sacrifices information of its high-resolution (HR) counterpart in favor of general utility, such as displaying or editing on smaller screens, low storage requirements, and fast transmission. Super-resolution attempts to recover the original HR copy from an LR input. However, the initial HR to LR transformation is commonly non-invertible and lossy \cite{nearestneighbor}. Thus, recovering the HR by estimating a Super Resolution (SR) analog is an ill-posed problem with the risk of a distorted output \cite{yang2014single}.

The classical interpolation methods for super-resolution only exploit local information and are thus incapable of generating commendable SR \cite{bicubic_2016}. While global image features extracted by the deep convolutional networks translated to a much-improved performance \cite{urban100,  haris2018dbpncvpr}, limited generalizability and distorted, SR remains a major concern \cite{dong2015srcnn}.

The landscape of super-resolution techniques had a significant breakthrough with the advent of the Generative Adversarial Network (GAN) \cite{goodfellow2020generative}. A super-resolution GAN \cite{srgan_cvpr2017} embraces the canonical two-player adversarial game between a generator $G$ and a discriminator $D$ with some minor modifications. Specifically, in a super-resolution task, $G$ attempts to map an LR input to an HR ground truth, generating an estimated SR in the process. The discriminator $D$ helps $G$ by providing adversarial feedback by distinguishing between an HR ground truth and $G$-generated SR. While GAN-based super-resolution offers generalizability through their generative power, they often have SR outputs that lose finer details or are plagued by artifacts \cite{checkerboardartifact_pixelshuffle, batchnorm_problem}.

\begin{figure}[!ht]
\centering
\includegraphics[width=0.6\textwidth]{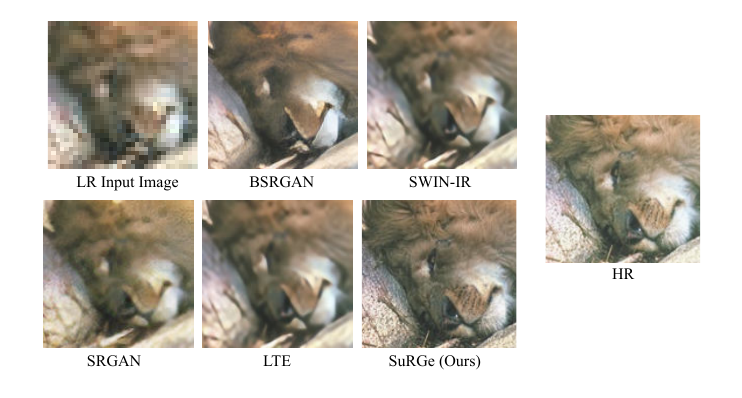}
 \caption{Visual comparison of 4x super-resolution outputs of the proposed SuRGe with SRGAN \cite{srgan_cvpr2017}, BSRGAN \cite{gu2019bsrgan}, SWIN-IR \cite{swiniriccvw2021}, and LTE \cite{lee2022lte}, given a low-resolution (LR) input image patch. SuRGe produces better super-resolution images with finer texture, color, and intricate details.}
\label{fig:intro}
\end{figure}

In this paper, we propose a GAN-based super-resolution method called \underline{Su}per-\underline{R}esolution \underline{Ge}nerator (SuRGe)\footnote{Code available at \url{https://github.com/Thecoder1012/SuRGe}}. In a super-resolution task, to generate a good quality SR image, it is necessary to consider both the low-level local features (for example, colors, textures, edges, etc.) and the high-level global ones (such as individual object shapes, relative positioning of objects and background, object orientation, etc.). As noted in \cite{resnet50, prelu}, higher-level global features are progressively captured by convolutional filters 
residing deeper in the network.
Taking inspiration from \cite{resnet50, ronneberger_unet} in the generator $G$ of the proposed SuRGe, we preserve the hierarchically complex features and dictate their flow through skip connections. However, skip connections may lead to under-utilized network capacity \cite{zhang2019revisiting} while a potential solution like DenseNet may be challenging to train \cite{zhang2021resnet} with limited data. Therefore, in SuRGe, we design a generator $G$ that judiciously uses the skip connections to conserve and propagate only a few selected features that are 
intuitively more useful
for improving the network's performance on super-resolution task (for example, carrying forward the low-level features to recover minute details after a potentially distortion inducing up-sampling step, in a spirit similar to that of UNet \cite{ronneberger_unet}). To adaptively combine the features coming from different network depths, we introduce mixing modules that operate in a learnable fashion.

We further focus on the fact that, ideally, the distributions of SR and HR should be identical. Thus, a loss function like Jensen–Shannon (JS) divergence that explicitly encourages minimizing the dissimilarity between the respective distributions of HR and SR helps train the generator $G$ in SuRGe. Moreover, in the ideal case, LR and SR should also preserve structural similarities, which consequently get reflected in their corresponding distributions. However, LR and SR reside in different metric spaces with potentially distinct dimensionality. Thus, to explicitly minimize the discrepancy between the respective distributions of LR and SR, we further utilize the Gromov Wasserstein (GW) distance \cite{memoli2011gromov} as an additional loss function in the generator $G$ of SuRGe. As per our knowledge, this is the first time the applicability of explicit divergence measures is explored in the context of GAN-based super-resolution techniques. We also employ a dynamically weighted convex combination strategy of the multiple losses in $G$ \cite{pmlr-v202-datta23a} of SuRGe. Furthermore, to prevent $G$ from mode collapsing, especially on the smaller training sets used in super-resolution \cite{DIV2K} we employ Wasserstein loss with gradient penalty (WGAN-GP) \cite{wgan-gp} to train the discriminator $D$ in SuRGe.

The primary contributions of our fully convolutional GAN-based SR method, SuRGe, are as follows. 
\begin{itemize}
    \item To the best of our knowledge, SuRGe is the first super-resolution model that introduces GW, a divergence between metric spaces of potentially different dimensions, to fuel the learning of generator $G$. This incorporation of the LR-SR relationship directly endows SuRGe with authentic super-resolution capabilities. 
    \item Moving away from pre-trained model-biased perceptual similarity, SuRGe takes JS divergence as an additional loss of generator $G$ (alongside adversarial and GW) while discriminator $D$ uses gradient penalized Wasserstein loss to improve the SR.
    \item We introduce a generator $G$ in SuRGe that efficiently employs skip connections to garner semantic information from different levels of feature representation and supports their adaptive mixing in a learnable fashion.
\end{itemize}
The effect of these critical improvements is evident in the motivational example in Figure \ref{fig:intro}, where, compared to four notable contenders, the SR obtained by SuRGe is richer in finer details and most closely matches the HR. Following a brief review of the existing deep super-resolution strategies using Convolutional Neural Networks and Transformers in Section \ref{sec:relatedWorks}, we detail the proposed methodology in Section \ref{sec:method}. In Section \ref{sec:experiment}, we show that the proposed SuRGe outperforms the current best by an average of 4.26\% and 5.72\%, respectively, in terms of PSNR \cite{PSNR} and SSIM \cite{wang2004ssim} on four common benchmarks for 4x super-resolution. Further, SuRGe supersedes the state-of-the-art by 17.59\% in terms of PSNR on six complex $4$x super-resolution datasets.

\section{Related Works}
\label{sec:relatedWorks}
Deep super-resolution models can be Convolutional Neural Networks (CNNs), GANs, and, more recently, transformers. The CNN-based methods are the first to employ deep networks for super-resolution \cite{dong2015srcnn}, using convolution maps through the image, followed by interpolation methodologies similar to a typical convolutional autoencoder. Though innovative for initial study and duly credited for a remarkable improvement over the traditional techniques, such CNN architectures suffer from poor generalizability and thus are domain-dependent \cite{urban100}. As a remedy, WDRN \cite{wdrntnnls2022} employs distinct wavelet features and their adaptive mixture for a better super-resolution performance. 

 The shortcomings of CNN can be addressed using GANs \cite{srgan_cvpr2017} with task-specific modifications. This route of research mainly diverges into three primary avenues. First, removing normalization and introducing dense blocks in the generator \cite{wang2018esrgan} indeed improves the SR image quality, although with a greater computational cost. Second, replacing dense networks with a residual backbone \cite{rakotonirina2020esrgan+} utilizing skip connections. Even though such networks are considerably easier to train, their full potential may not be realized without carefully curating the skip connections and feature mixing that best aid the super-resolution task. Third, additional preprocessing, such as blurring and specialized noise injection \cite{wang2021realesrgan, gu2019bsrgan}, produces augmentations that are close to real-life scenarios and can lead to more enhanced SR output. Unfortunately, this also purposefully distorts the input distribution that may, in turn, sacrifice the clarity and details. On the other hand, considering perceptual similarity is introduced in \cite{li2022bybygan} that directly depends upon the generalization of an external pre-trained network for optimizing the generator. In summary, such methods typically suffer from loss of minute details \cite{gu2019bsrgan} or distorted boundaries \cite{li2022bybygan}. Moreover, all of these methods perform the required up-scaling at once at the end of the network. Thus, the network cannot mitigate any possible distortion during the drastic up-scaling. Furthermore, the ever-improving GAN variants remain mostly unexplored in the context of super-resolution.

 SAN \cite{sancvpr2019} introduces channel attention in CNN to open the gate for transformer networks in super-resolution. In support of a better performance, transformers not only manage to adaptively mix diversely informative features through attention \cite{swiniriccvw2021, zhang2022swinfir} but also mitigate SR output distortions using layer normalization. In \cite{chu2022nafssr}, the idea of cross-attention is proposed, which was later improved in \cite{chen2205hat-l} to mitigate the adverse impacts of uncontrolled mixing of distinct features. The DAT \cite{daticcv2023} introduces a novel transformer model that aggregates features through inter-block spatial and intra-block channel attentions. In essence, they introduce the Adaptive Interaction Module (AIM) and the Spatial-Gate Feed-Forward network (SGFN) for a tailored feature aggregation at different levels. Later, SR-Former \cite{srformericcv2023} attempts to improve \cite{daticcv2023} by focusing on Permuted Self Attention (PSA) for a more balanced approach towards feature aggregation through channel and spatial attention. 
 
Recently, \cite{tran2024cpat} addresses shortcomings of local-window attention by introducing Channel-Partitioned Windowed Attention (CPAT), expanding windows along Spatial-Frequency Interaction (SFIM) to capture global context and intricate details. Following this, DRCT \cite{hsu2024drct} proposes Dense-Residual Connections to stabilize the gradient flow across layers while preserving the spatial information despite ``information bottleneck''. In a somewhat similar line, HMA-Net \cite{chu2024hmanet} mixes different attention types such as grid and residual transformer blocks, which essentially incorporate hierarchical abstraction (Hi-IR) for richer pixel-to-semantic features. MaIR \cite{li2024mair} proposes a multi-directional Sequence Shuffle Attention (SSA) using Nested S-shaped Scanning to preserve 2D locality by scanning images in nested stripes. In essence, the focus remains on identifying effective feature mixing alongwith judicious preservation and propagation of information that best aids the super-resolution task.

\section{Proposed Method}
\label{sec:method}
Typically, the image matrix (or vector if flattened) resides in lower-dimensional ambient space $\mathcal{M}_1$ \cite{pope2021the} i.e., a low-resolution (LR) image $\mathbf{x} \in \mathcal{M}_1$. Thus, a higher-resolution (HR) version of $\mathbf{x}$ exists as $\mathbf{y} \in \mathcal{M}_{2}$ that improves definition. The estimated form of $\mathbf{y}$ is known as the super-resolution output SR. Commonly $\mathcal{M}_{1} \subset \mathbb{R}^{whc}$ and $\mathcal{M}_{2} \subset \mathbb{R}^{w'h'c}$, where $w'=wr$, $h'=hr$, and $r \in \mathbb{Z}^{+}$ is the multiplicative scaling factor \cite{wang2018prosrl} denoting the extent of magnification from LR to HR (or SR). A GAN-based super-resolution method given input $\mathbf{x}$ searches for a generator $G \in \{G_{\theta}:\mathcal{M}_1 \rightarrow \mathcal{M}_2| \theta \in \Theta\}$ that minimizes the discrepancy between SR $G(\mathbf{x})$ and its HR analog $\mathbf{y}$. A discriminator $D$ guides $G$ by providing feedback through distinguishing $\mathbf{y}$ and $G(\mathbf{x})$. We further denote the distributions of LR, HR, and SR as $p_{\mathbf{x}}$, $p_{\mathbf{y}}$, and $p_{G(\mathbf{x})}$ respectively.

\begin{figure}[!ht]
    \centering
    \includegraphics[width=.5\textwidth]{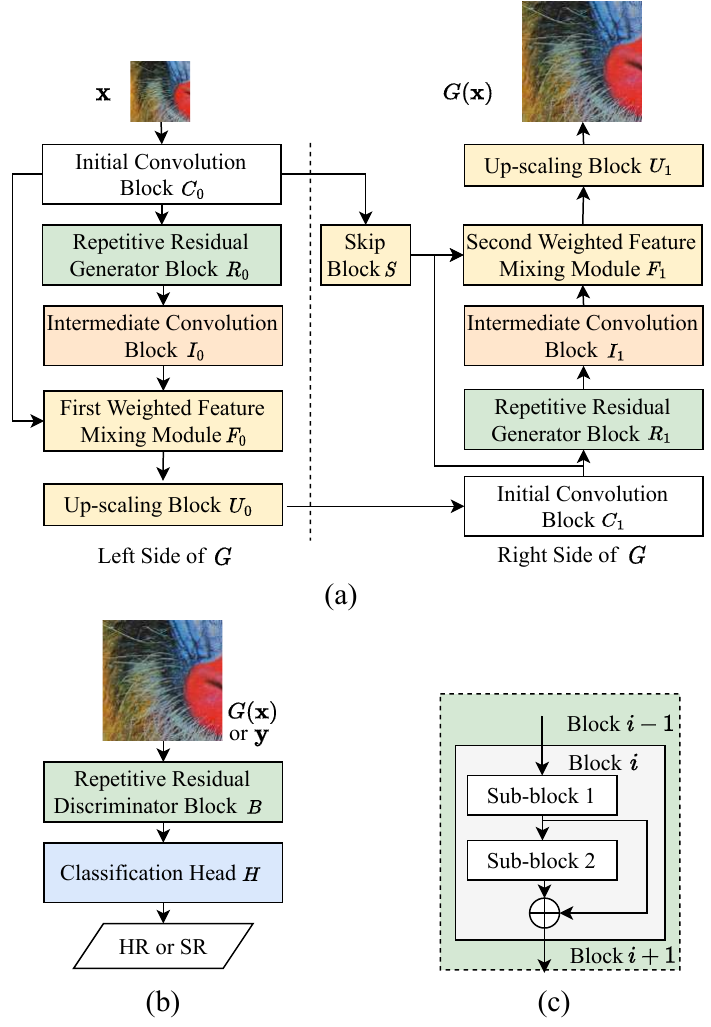}
    \caption{The schematic of SuRGe demonstrates two of its main components in (a) the generator $G$ and (b) the discriminator $D$. Moreover, in (c), we detail the structure of our sub-network Repetitive Residual Block, used in $G$ and $D$. $G$ takes a LR image $\mathbf{x}$ and generates a 4x up-scaled SR image $G(\mathbf{x})$. $D$ guides $G$ by distinguishing an input between HR ground truth $\mathbf{y}$ and SR $G(\mathbf{x})$. Further details on network design can be found in the \ref{app:sec:networkDetails}.}
    \label{fig:surge schematic}
\end{figure}

\subsection{The Architecture of $G$}
\label{sec:generator}
The generator $G$ aims to recover $\mathbf{y}$ from $\mathbf{x}$ under the commonly used constraint of $r=4$ i.e. through 4x super-resolution \cite{dong2015srcnn, srgan_cvpr2017, wang2018esrgan, swiniriccvw2021, chu2022nafssr}. Unlike popular practice \cite{swiniriccvw2021, chu2022nafssr} of performing 4x up-scaling in one shot at the end, $G$ in SuRGe performs the same in two steps, i.e., a 2x up-scaling (see Figure \ref{fig:surge schematic}) at the end of each half. This way, the right half of $G$ can mitigate the possible abrupt distortions of the feature space due to the first 2x up-scaling and reuse features from the left half (through skip connections) to recover corrupted information. We demonstrate $G$ in Figure \ref{fig:surge schematic}, highlighting the key components while detailing them individually in the following. 

The initial convolution block $(C_{0})$ extracts the low-level features using larger kernels with half-padding, providing two benefits. \textbf{(1)} Repetitive information and its variation over a larger region can be better captured \cite{color_bianco_2015}. \textbf{(2)} Possible distortions near the image boundaries can be avoided \cite{dong2015srcnn}. Moreover, we use parametric ReLU to allow the distinct layers to have different non-linearity for better conservation of low-level features. Further, we discard normalization to avoid information loss through regularization and scaling.

The repetitive residual generator block $(R_{0})$ focuses on extracting high-level intricate features using smaller kernels. This contains $n_{G}$ residual blocks \cite{resnet50}, each having two sub-blocks. The first sub-block alone uses parametric ReLU activation, while both employ batch normalization to induce regularization and limit covariance shift. The outputs of the two sub-blocks are added through a skip connection and passed to the next residual block. The inter-sub-block skip connection ensures that the features after each convolution and normalization at least retain the extracted information, if not able to enrich it further.

The $n_{G}$-th block of $R_{0}$ sums the outputs of its two sub-blocks. This distorted feature space, similar to a typical ResNet, must be further stabilized before being processed in the next stage. However, the common remedy of average pooling fails in super-resolution because neither is such a kernel learned nor does the down-scaling go along with the task objective. Therefore, the intermediate convolution block $(I_0)$ is added to stabilize the output of $R_{0}$ by additional convolutions with batch normalization.

At the outputs of $C_{0}$ and $I_{0}$, we respectively have low and high-level features. Thus, we use the first weighted feature mixing module $F_{0}$ to combine these two features before up-scaling. $F_{0}$ performs a simple convex combination as:
\begin{equation}
    F_{0} = w^{(F_{0})}_{1}C_{0} + w^{(F_{0})}_{2}I_{0}(R_{0}(C_{0})),
\end{equation}
 where $w^{(F_{0})}_{1}, w^{(F_{0})}_{2} > 0$ and $w^{(F_{0})}_{1}+w^{(F_{0})}_{2}=1$. We learn both of $w^{(F_{0})}_{1}$ and $w^{(F_{0})}_{2}$ as parameters of $G$ while the convexity constraint is ensured by passing the weights through a Softmax activation. 

\begin{figure}[!t]
    \centering		
    \includegraphics[width=.6\textwidth]{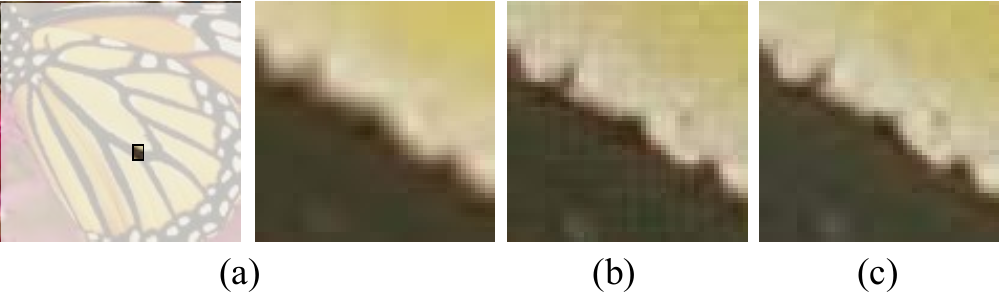}
    \caption{In comparison to HR patch (a) of a butterfly image, the checkerboard pattern introduced by PixelShuffle \cite{checkerboardartifact_pixelshuffle} is apparent in (b). Nearest neighbor up-scaling in SuRGe generates clean $G(\mathbf{x})$ as evident from our result in (c).}
    \label{fig:checkerboard}
\end{figure}

The output of $F_{0}$ is first stabilized with convolution and then passed to $U_{0}$ for 2x up-scaling. The commonly used up-scaling techniques, such as PixelShuffle and transposed convolutions, though effective otherwise, are likely to distort SR as in Figure \ref{fig:checkerboard}. This is because the overlapped kernels may introduce uneven convolution that results in higher frequency color patterns, like a checkerboard in the border pixels of the kernel mapping. Hence, we employ the interpolation-based 
nearest neighbor method
\cite{nearestneighbor} for up-scaling.

 Except for $F_{1}$, the rest of the right half of $G$ is identical to the left. The right or second half of $G$ starts with an initial convolution layer $C_{1}$ that stabilizes the $U_{0}$ output. In an attempt to recover any lost or corrupted information at this stage, we add the output of $C_{1}$ with the 2x up-sampled output of $C_{0}$ after passing it through a skip block $S$. The sum is then propagated through $R_{1}$ and $I_{1}$ layers. $F_{1}$ takes the following three inputs: \textbf{(1)} $C_{0}$ output 2x up-scaled by a skip block $S$ having a structure similar to $U \circ C$ that recovers the low-level features at the end of the network. \textbf{(2)} The output of $I_{1}$. \textbf{(3)} The output of $C_{1}$. Similar to $F_{0}$, here also we perform a convex combination of the three inputs as follows:
\begin{equation}
    F_{1} = w^{(F_{1})}_{1}I_{1} + w^{(F_{1})}_{2}C_{1} + w^{(F_{1})}_{3}S(C_{0}),
\end{equation}
 where the three weights $w^{(F_{1})}_{1}$, $w^{(F_{1})}_{2}$, and $w^{(F_{1})}_{3}$ are constrained with convexity similarly to their counterparts in $F_{0}$ and are thus learned in the same way. The output of $F_{1}$ is 2x up-scaled by $U_{1}$, and stabilized by further convolutions (without batch normalization or activation to avoid distortion) to produce the SR output $G(\mathbf{x})$.

\subsection{The Architecture of $D$}
\label{sec:discriminator}
 As demonstrated in Figure \ref{fig:surge schematic}, $D$ has two main components: a sub-network called Repetitive Residual Discriminator Block $B$ and a classification head $H$. The structure of $B$ mostly follows $R$ as it contains $n_{D}$ residual blocks, each with two sub-blocks connected by an inter-sub-block skip connection. Maintaining near structural similarity between $G$ and $D$ enables the same input to likely have close embeddings in the learned feature space. Thus, $D$ can easily identify a deviation of $G(\mathbf{x})$ from $\mathbf{y}$ and improve $G$ through a more useful feedback.

 There are three key differences between $R$ and $B$. \textbf{(1)} $B$ uses LeakyReLU activation to prevent sparse or scattered gradients \cite{goodfellow2020generative}. \textbf{(2)} $B$ employs Pixel normalization \cite{karras2017progan}, as batch normalization is known to cause quality issues in a super-resolution task when used in $D$ \cite{batchnorm_problem, wang2018esrgan}. Furthermore, the specific WGAN-GP nature of our training regime prohibits the discriminator from using batch normalization. \textbf{(3)} The number of filters and the convolution kernel size are gradually increased over the residual blocks. This not only improves the balanced capture of low-level and high-level information but also prevents over-fitting by removing bias to a particular kernel size.

 The classification head $H$ first performs an adaptive average pooling on the output of $B$. The pooled features are then flattened and passed through dense layers with LeakyReLU activation. 
 The final dense layer maps the features to a single node but does not apply a non-linear activation to it, respecting a typical WGAN-GP strategy.

\subsection{Loss functions of SuRGe}
\label{sec:lossFunctions}
 SuRGe, embodying the GAN philosophy, has tailor-made losses for the generator $G$ and the discriminator $D$.

 \subsubsection{Loss for generator $G$:} To receive guidance from $D$, $G$ utilizes a traditional adversarial loss $\mathcal{L}^{G}_{a}$ defined as:
\begin{equation}
    \label{eq:advloss}
    \mathcal{L}^{G}_{a} = - \sum\nolimits_{\mathbf{x} \in N} \log{D(G(\mathbf{x}))},
\end{equation}
 where $N$ is a training batch. 

 The classical $\mathcal{L}^{G}_{a}$, though necessary, is not sufficient for maintaining the desired perceptual quality of the SR output when deployed alone. A common remedy \cite{srgan_cvpr2017, rakotonirina2020esrgan+} is to additionally minimize the discrepancy between HR and SR in the embedding space of a pre-trained deep network that is likely capable of expressing perceptual information. Evidently, the efficacy of such a loss is reliant on the quality and generalizability of the pre-trained embedding space \cite{wang2018esrgan}. However, $p_{\mathbf{y}}$ and $p_{G(\mathbf{x})}$, respectively the distributions of HR and SR, in practice are supported on the same ambient space. Hence, directly minimizing their divergence using a symmetric measure like JS motivates $G(\mathbf{x})$ to resemble $\mathbf{y}$:
\begin{equation}
    \label{eq:jsLoss}
    \mathcal{L}^{G}_{\textrm{JS}} = \frac{1}{2}\mathbb{E}_{p_{\mathbf{y}}}\left[\log(p_{\mathbf{y}})-\log\left(\frac{(p_{\mathbf{y}} + p_{G(\mathbf{x})})}{2}\right)\right] + \frac{1}{2}\mathbb{E}_{p_{G(\mathbf{x})}}\left[\log(p_{G(\mathbf{x})}) - \log\left(\frac{1}{(p_{G(\mathbf{x})} + p_{\mathbf{y}})}{2}\right)\right].
\end{equation}

 As the name suggests, at the heart of the super-resolution problem lies the task of learning a meaningful transformation $G$ that refines LR images visually. The optimization, however, is constrained based on the need to preserve semantic features. Such information in a set of samples is stored not only in coordinate entries of the vectors but also in their local geometry. As such, a generative model becomes a true SR architecture based on its capacity to keep the metric measure spaces corresponding to $p_{\mathbf{x}}$ and $p_{G(\mathbf{x})}$ near-isometric. The divergence that enables penalizing the deviation from such an ideal scenario is GW. Thus, in SuRGe, we integrate the GW loss in training $G$:      
\begin{equation}
    \label{eq:gwLoss}
    \mathcal{L}^{G}_{\textrm{GW}} = \min_{\gamma \in \Gamma} \int |d_{1}(\mathbf{x},\mathbf{\Tilde{x}}) - d_{2}(\mathbf{z},\mathbf{\Tilde{z}})|^{2} d\gamma(\mathbf{x},\mathbf{z}) d\gamma(\mathbf{\Tilde{x}},\mathbf{\Tilde{z}}),
\end{equation}
 where $\Gamma$ is the set of couplings between distributions $p_{\mathbf{x}}$ and $p_{G(\mathbf{x})}$, while $\mathbf{x}, \bar{\mathbf{x}} \sim p_{\mathbf{x}}$, and $\mathbf{z}, \bar{\mathbf{z}} \sim p_{G(\mathbf{x})}$. Also, $d_{1}, d_{2}$ are the metrics on the spaces $\mathcal{M}_1$ and $\mathcal{M}_2$ respectively. {Given that our applications only consider images, both the metric spaces $(\mathcal{M}_1, d_{1})$ and $(\mathcal{M}_2, d_{2})$ are Euclidean of amenable ambient dimensions, endowed with $L^2$ norms. Our implementation, in finding the optimal coupling $\gamma \in \Gamma$, follows the optimization regime of \cite{Nakagawa2023, chakrabarty2024robust}. Since the target coupling is characterized using $G_{\theta}$, we calculate the empirical GW distance directly instead of updating $\gamma$ in a nested optimization scheme. This in turn enables us to compute the GW loss as the mean squared distance between the Euclidean cost matrices that takes a total of $\mathcal{O}(N^{2}r^{3}whc)$. In our particular case of $r=4$ we get $\mathcal{O}(64N^{2}whc)$ \cite{Nakagawa2023, chakrabarty2024robust}.}

 Tuning a set of static weights to combine the three loss components in $\mathcal{L}_{G}$ is not only tedious but also inefficient due to being oblivious to dynamic training situations. Learning the weights as network parameters may also bias the training towards a particular component. Thus, we employ a convex combination of the three loss components where the weights are dynamically calculated \cite{pmlr-v202-datta23a}. Specifically, the values of the three loss components are passed through Softmax. As such, the dynamically assigned weight to a loss component depends on its value such that at any point in time, the weights adjust according to the values for preventing the dominance of one over the others in the combined $\mathcal{L}^{G}$. In essence, at each iteration of training:
\begin{equation}
    \label{eq:dynLoss}
    \mathcal{L}^{G} = w_{a}\mathcal{L}_{a}^{G} + w_{\textrm{JS}}\mathcal{L}_{\textrm{JS}}^{G} + w_{\textrm{GW}}\mathcal{L}_{\textrm{GW}}^{G}, \; \text{where} \; w_{(\cdot)} = \frac{\exp({\mathcal{L}_{(\cdot)}^{G}})}{(\exp({\mathcal{L}_{a}^{G}})+\exp({\mathcal{L}_{\textrm{JS}}^{G}})+\exp({\mathcal{L}_{\textrm{GW}}^{G}}))}.
\end{equation}
Now $w_{(\cdot)}$ can be $w_{a}$, $w_{\textrm{JS}}$ and $w_{\textrm{GW}}$ while $\mathcal{L}_{(\cdot)}^{G}$ is respectively set to $\mathcal{L}_{a}^{G}$, $\mathcal{L}_{\textrm{JS}}^{G}$, and $\mathcal{L}_{\textrm{GW}}^{G}$.

 \subsubsection{Loss for discriminator $D$:} We draw inspiration from WGANs' promise of improving generation quality by deploying the Wasserstein-$1$ distance (WD) to distinguish between `real' and `fake' samples. The underlying class of critic functions (them being $k$-Lipschitz continuous) additionally mollifies mode collapse \cite{arjovsky2017wasserstein}. However, maintaining $k$-Lipschitz continuity during training is difficult as it requires limiting the gradients of $D$. WGAN achieves this by a weight-clipping heuristic that sacrifices complexity. As a better alternative, WGAN-GP puts a constraint on the gradient itself that can be expressed as a regularizer called the gradient penalty. Thus, $\mathcal{L}^{D}$ can be written as follows: 
\begin{equation}
    \label{eqn:DLoss}
    \mathcal{L}^{D} = \mathbb{E}_{p_{\mathbf{x}}} D(G(\mathbf{x})) -\mathbb{E}_{p_{\mathbf{y}}} D(\mathbf{y})
    +\lambda \mathbb{E}(||\nabla_{\hat{\mathbf{x}}}D(\hat{\mathbf{x}})||_{2} -1)^{2},
\end{equation}
 where $\hat{\mathbf{x}} = \epsilon\mathbf{y} + (1-\epsilon)G(\mathbf{x})$, and $\epsilon \sim \textrm{Uniform}(0,1)$. 

\begin{figure}[!ht]
    \centering
    \includegraphics[width=.6\textwidth]{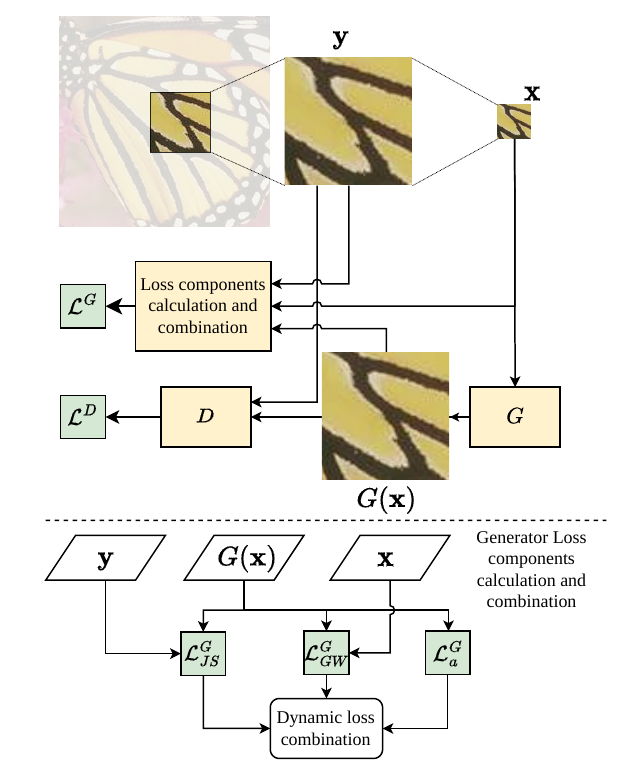}	
    \caption{We extract patch from HR as $\mathbf{y}$ and down-scale it to LR input $\mathbf{x}$. The $\mathbf{x}$ is fed to $G$ to obtain the 4x SR $G(\mathbf{x})$. The $G(\mathbf{x})$ is used for $\mathcal{L}^{G}_{a}$ using equation (\ref{eq:advloss}), $G(\mathbf{x})$ and $\mathbf{y}$ together is used for $\mathcal{L}^{G}_{\textrm{JS}}$ using equation (\ref{eq:jsLoss}), and $\mathbf{x}$ with $G(\mathbf{x})$ find $\mathcal{L}^{G}_{\textrm{GW}}$ using equation (\ref{eq:gwLoss}). We take the Softmax-based dynamic convex combination of $\mathcal{L}^{G}_{a}$, $\mathcal{L}^{G}_{\textrm{JS}}$, and $\mathcal{L}^{G}_{\textrm{GW}}$ as per equation (\ref{eq:dynLoss}) to find $\mathcal{L}^{G}$ to update $G$. For updating $D$, we use $\mathbf{y}$ and $G(\mathbf{x})$ to calculate $\mathcal{L}^{D}$ using equation (\ref{eqn:DLoss}).}
  \label{fig:workflow}
\end{figure}

\subsection{Putting it all together}
 The workflow of SuRGe is illustrated in Figure \ref{fig:workflow} while the algorithm is described in Algorithm \ref{alg:surge} in \ref{app:sec:algo}. We follow a patch-based training \cite{wang2018prosrl, srgan_cvpr2017} in SuRGe. The idea is to extract a $256 \times 256$ overlapped patch of the HR ground truth as $\mathbf{y}$ and 4x bi-cubic down-scale the same to $64 \times 64$ to get the corresponding LR as $\mathbf{x}$. The training strategy of SuRGe is similar to a vanilla GAN \cite{goodfellow2020generative}. Thus, $G$ and $D$ are alternatively updated with the gradients of the respective $\mathcal{L}^{G}$ and $\mathcal{L}^{D}$ loss.

\section{Experiments}
\label{sec:experiment}
 \subsection{Experimental Protocol} Following standard practice, we train SuRGe on DIV2K \cite{DIV2K} dataset using the 800 training examples. We test SuRGe on four popular benchmarks namely Set5 \cite{set5}, Set14 \cite{set14}, BSD100 \cite{bsd100}, and Urban100 \cite{urban100} along with six additional datasets viz. Kitti2012, Kitti2015 \cite{kittidataset}, Middlebury \cite{middlebury}, PIRM \cite{pirm}, OST300 \cite{wang2018ost300} and MANGA109 \cite{manga109}. The details of network architecture, datasets, pre-processing of input, the hyper-parameter choices, and their tuning with grid search are provided respectively in \ref{app:sec:networkDetails}, \ref{app:sec:dataDescription}, and \ref{app:sec:networkSearch}. We use PSNR \cite{PSNR} and SSIM \cite{wang2004ssim} to measure the performances of all the methods, both of which are described in \ref{app:sec:metrics}. The code for SuRGe is currently provided at the public GitHub repository \url{https://github.com/Thecoder1012/SuRGe} for ease of access and result reproduction. 

 \begin{table*}[!ht]
    \centering
    \caption{Ablation study of SuRGe on the BSD100 dataset in terms of PSNR and SSIM. The gradual improvement in performance with the progressive addition of key components through seven intermediate models ($V_{0}-V_{4}$ including two variants for each of $V_{3}$ and $V_{4}$) validates their importance in SuRGe. The best result is boldfaced.}
    \label{tab:ablation}
    \scriptsize
    \vspace{5pt}
    \resizebox{\textwidth}{!}{
    \begin{threeparttable}
    \begin{tabular}{l|cccc|cccc|cc|ccc|cc}
         \toprule
            Model & $D_{\textrm{VGG}}$ & $D_{\textrm{RES}}$ & $D_{3\textrm{K}}$ & $D_{\textrm{IK}}$ & $\mathcal{L}^{G}_{p}$ & $\mathcal{L}^{G}_{\textrm{JS}}$ & $\mathcal{L}^{G}_{\textrm{GW}}$ & DS & $G_{\textrm{CC}}$ & $G_{F_{0,1}}$ & $\mathcal{W}^{G}_{t}$ & $\mathcal{W}^{G}_{l}$ & $\mathcal{W}^{G}_{dw}$ & PSNR$^{1}$ & SSIM$^{1}$ \\ \midrule
            $V_{0}$ & \checkmark & & \checkmark & & \checkmark & & & & \checkmark & & \checkmark & & & $29.61$ & $0.76$\\
            $V_{1}$ & & \checkmark & \checkmark & & \checkmark & & & & \checkmark & & \checkmark & & & $30.04$ & $0.81$ \\
            $V_{2}$ & & \checkmark &  & \checkmark & \checkmark & & & & \checkmark & & \checkmark & & & $30.14$ & $0.81$ \\ \midrule
            $V_{3,RN}$ & & \checkmark &  & \checkmark & & \checkmark & & RN & & \checkmark & \checkmark & & & $29.85$ & $0.83$ \\
            $V_{3,FI}$ & & \checkmark &  & \checkmark & & \checkmark & & FI & & \checkmark & \checkmark & & & $30.99$ & $0.84$ \\ \midrule
            $V_{4,RN}$ & & \checkmark &  & \checkmark & & \checkmark & \checkmark & RN & & \checkmark & & \checkmark & & $30.16$ & $0.83$ \\
            $V_{4,FI}$ & & \checkmark &  & \checkmark & & \checkmark & \checkmark & FI & & \checkmark & & \checkmark & & $31.29$ & $0.86$ \\
            \midrule
            SuRGe & & \checkmark &  & \checkmark & & \checkmark & \checkmark & FI & & \checkmark & & & \checkmark & \textbf{31.52} & \textbf{0.87} \\
            \bottomrule
    \end{tabular}
    \begin{tablenotes}
        \item $D_{\textrm{VGG}}$: $D$ with VGG-type backbone. $D_{\textrm{RES}}$: $D$ with ResNet-type network. $D_{3\textrm{K}}$: Kernel size is set to 3 in $D$. $D_{\textrm{IK}}$: $D$ with incremental kernel size.  $\mathcal{L}^{G}_{p}$: Perceptual similarity calculated with ResNet-50 is used as a loss. $\mathcal{L}^{G}_{\textrm{JS}}$: Jensen-Shannon Divergence calculated between SR and HR image batch. $\mathcal{L}^{G}_{\textrm{GW}}$: Gromov-Wasserstein Loss calculated between LR and SR image batch. DS: Space used for divergence measures, can be ResNet-50 (RN) or Flattened Image (FI). $G_{\textrm{CC}}$: During combinations, the feature from the main path is taken entirely while the skip connection weights are manually tuned. $G_{F_{0,1}}$: $G$ using the $F_{0,1}$ in SuRGe.  $\mathcal{W}^{G}_{t}$: Summing $\mathcal{L}^{G}_{a}$, $\mathcal{L}^{G}_{\textrm{JS}}$, $\mathcal{L}^{G}_{\textrm{GW}}$ in the Generator Loss $\mathcal{L}^{G}$. $\mathcal{W}^{G}_{l}$: Mixing $\mathcal{L}^{G}_{a}$, $\mathcal{L}^{G}_{\textrm{JS}}$, $\mathcal{L}^{G}_{\textrm{GW}}$ in the Generator Loss $\mathcal{L}^{G}$ with weights learned as parameters of the network. $\mathcal{W}^{G}_{dw}$: Mixing $\mathcal{L}^{G}_{a}$, $\mathcal{L}^{G}_{\textrm{JS}}$, $\mathcal{L}^{G}_{\textrm{GW}}$ with the dynamic weighting in the Generator Loss $\mathcal{L}^{G}$. $^{1}$: Increment indicates improvement.
    \end{tablenotes}
    \end{threeparttable}}
\end{table*}

\begin{figure}[!ht]
	\centering
		\includegraphics[width=0.6\textwidth]{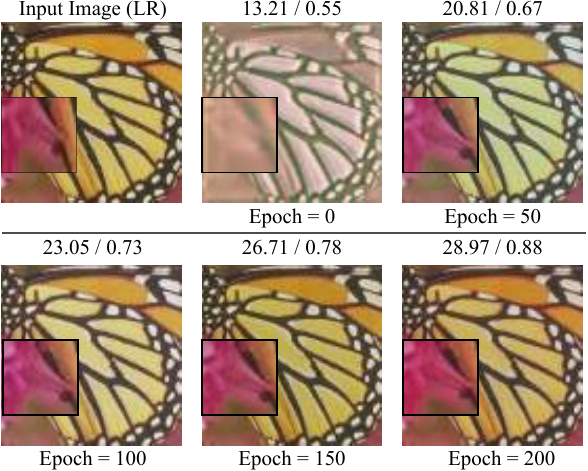}	
		\caption{We show the generated SR patch for a butterfly (Set5) test instance, along with the metrics (on top as PSNR/SSIM) in the intervals of every 50 training epochs of SuRGe. The gradual improvement in the SR output of SuRGe is apparent with the progress in training. }
  \label{fig:epoch_wise_img}
\end{figure}

 \subsection{Ablation study} We start with an ablation study of the five critical components in SuRGe, namely the choice of backbone in $D$, the kernel size for convolution in $D$, the choice of loss functions in $G$, the combination strategy of $F_{0,1}$ in $G$, and the loss function in $D$. Moreover, for the choice of loss functions in $G$, we additionally consider the space where the two divergence measures can be calculated, giving us a total of seven initial variants. Table \ref{tab:ablation} shows that over the aforementioned seven intermediate models on the BSD100 dataset, the performance gradually improves in terms of PSNR and SSIM with better choices for the components. The best performance is achieved when all the components act in harmony, validating their importance in SuRGe. Specifically, we start with comparing the usefulness of ResNet over VGG in $D$ between $V_{0}$ and $V_{1}$. Then, in $V_{2}$ we apply an incremental kernel in the $D$ instead of the constant one in $V_{1}$. In the following $V_{3}-V_{4}$, we move away from the traditional perceptual loss used in super-resolution and successively introduce the JS and GW divergence measures with the adversarial loss in $G$. Furthermore, in $V_{3}-V_{4}$ we implement the $F_{0,1}$ modules. Between $V_{3}$ and $V_{4}$ variants, the strategy of directly summing the loss components in $G$ is first upgraded with learnable coefficients that are finally replaced by dynamic weighting in SuRGe. Previous uses of GW \cite{bunne2019learning} argued in favor of representations obtained from a pre-trained deep network to limit the dimensions and improve stability. However, the particular task of super-resolution may benefit from the flattened images as they mitigate the risk of unregulated information alteration through feature extraction. We empirically confirm this through a comparison of $V_{3,RN}$ and $V_{4,RN}$ passing image features extracted from a pre-trained ResNet-50 against $V_{3,FI}$ and $V_{4,FI}$ feeding flattened images to the $\mathcal{L}^{G}_{\textrm{JS}}$ and $\mathcal{L}^{G}_{\textrm{GW}}$. Our experiment shows that using flattened images gives a performance boost compared to the ResNet-50 embedding space. As an extension, in Figure \ref{fig:epoch_wise_img}, we further show that the SR output generated by SuRGe progressively improves over training.

\begin{figure}[!ht]
	\centering
		\includegraphics[width=0.99\textwidth]{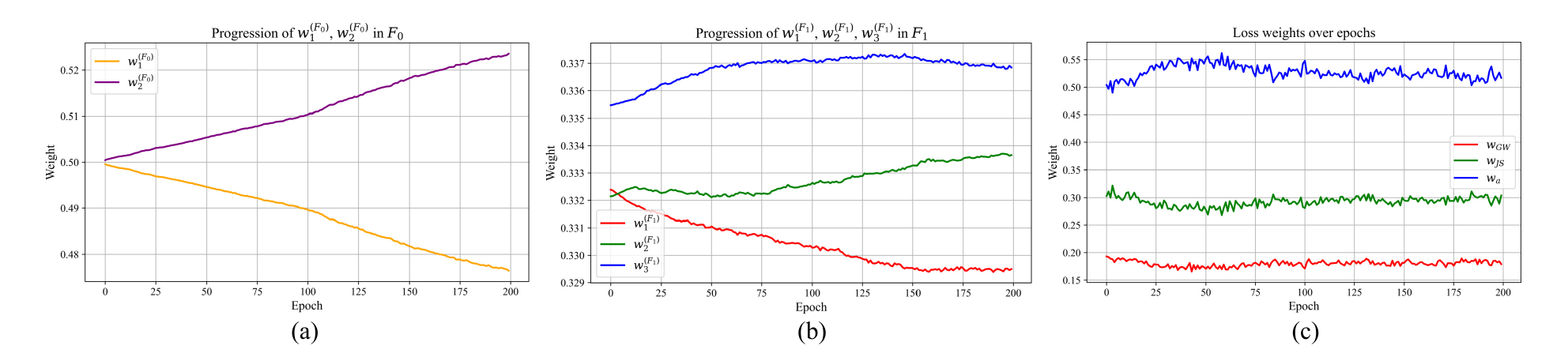}	
		\caption{In SuRGe, we have three weighting components, namely the learnable weights associated with the feature mixing modules $F_{0}$ and $F_{1}$, and the SoftMax-based dynamic loss-weighting to obtain the aggregated loss for the generator $G$. In this picture, we show how such weights actually converge over the training epoch. (a) The two learnable weights for the mixing module $F_{0}$. (b) The three learnable weights associated with the mixing module $F_{1}$. (c) The loss-value driven temperature-aided Softmax-based dynamic weights corresponding to the three losses in the generator $G$.}
  \label{fig:epoch_wise_weight}
\end{figure}

In a follow-up study, we take a closer look at the progression and convergence of the different weighting components associated with SuRGe. In the proposed method, we have two sets of learnable weights that respectively control the feature mixing in $F_{0}$ and $F_{1}$. These weights are learned as the parameters of $G$, i.e., tuned by the 
gradient descent
over the generator loss. Thus, they are expected to converge along with the training of the model. On the other hand, the three loss components of the generator $G$ are aggregated by a dynamic weighting scheme. In essence, a Softmax over the loss values (smoothed by a suitable temperature hyper-parameter) controls the extent to which a particular loss should be focused during a certain stage of training. Unlike the previous set of learnable weights, these ones are more likely to remain somewhat volatile, although in a limited range. To elaborate, the volatility indicates that the weights can adapt and shift focus on a non-converging loss as needed. In contrast, the limited range of such weights dictates that large changes in a loss are discouraged, i.e., no one loss is allowed to diverge or converge to an exceptional extent that may disrupt the optimization. To validate these intuitions, we plot the epoch-wise weights obtained during training in Figure \ref{fig:epoch_wise_weight}. From Figures \ref{fig:epoch_wise_weight}-(a) and \ref{fig:epoch_wise_weight}-(b), we can see that the learnable feature mixing weights indeed converge with the increasing training of the model. The convergence is more apparent in the case of $F_{1}$, where the three weights mostly become stationary after 150 epochs. For $F_{0}$, the convergence is slightly delayed, and the rate of change starts to slow down after about 180 epochs. Now, from Figure \ref{fig:epoch_wise_weight}-(c), we can see that the dynamic loss weighting scheme is also demonstrating the expected behavior that is altering at a higher rate while being bounded in a small range only after 25 epochs. Therefore, the weighting components in SuRGe and their update strategies indeed play a pivotal role in its success.

\begin{table}[!ht]
\scriptsize
    \centering
    \caption{Performance comparison of SuRGe in terms of PSNR and SSIM on four benchmarks against notable competitors.}
    \label{tab:quant_eval}
    \scriptsize
    \vspace{5pt}
    \begin{tabular}{llcccccccc}
    \toprule
    \multirow{2}{*}{Method} & \multirow{2}{*}{Strategy$^{2}$} & \multicolumn{2}{c}{Set5$^{1}$} & \multicolumn{2}{c}{Set14$^{1}$} & \multicolumn{2}{c}{BSD100$^{1}$} & \multicolumn{2}{c}{Urban100$^{1}$} \\ 
    \cmidrule{3-10} 
        & & PSNR$^{3}$ & SSIM$^{3}$ & PSNR & SSIM & PSNR & SSIM & PSNR & SSIM \\
        \midrule
        SRCNN & CNN & 30.49 & 0.86 & 27.50 & 0.75 & 26.91 & 0.71 & 24.53 & 0.72 \\
        SelfExSR & CNN &  \texttt{-{}-} &  \texttt{-{}-} &  \texttt{-{}-} &  \texttt{-{}-} & 26.80 & 0.71 & 24.67 & 0.73 \\ 
        DBPN-RES & CNN & 32.65 & \underline{0.90} & 29.03 & 0.79 & 27.82 & 0.74 & 27.08 & 0.81 \\ \midrule
        SRGAN & GAN & 29.40 & 0.85 & 26.02 & 0.74 & 23.16 & 0.67 &  \texttt{-{}-} &  \texttt{-{}-} \\
        ProSR-L & GAN &  \texttt{-{}-} &  \texttt{-{}-} & 28.94 &  \texttt{-{}-} & 27.68 &  \texttt{-{}-} & 26.74 &  \texttt{-{}-} \\
        ESRGAN & GAN & 32.73 & 0.90 & 28.99 & 0.79 & 27.85 & 0.75 & 27.03 & 0.82 \\
        RankSRGAN & GAN &  \texttt{-{}-} &  \texttt{-{}-} & 26.57 & 0.65 & 25.57 & 0.65 &  \texttt{-{}-} &  \texttt{-{}-} \\ 
        Beby-GAN & GAN & 27.82 & 0.80 & 26.96 & 0.73 & 25.81 & 0.68 & 25.72 & 0.77 \\
        Gram-GAN & GAN & 27.97 & 0.80 & 26.96 & 0.77 & 26.32 & 0.74 & 25.89 & 0.77 \\ \midrule
        SAN & CNNA & 32.70 & \underline{0.90} & 29.05 & 0.79 & 27.86 & 0.75 & 27.23 & 0.82 \\
        WRAN & CNNA & 28.60 & \underline{0.90} & 28.60 & 0.79 & 27.71 & 0.74 & 26.74 & 0.80 \\ \midrule
        SwinIR & TRAN & 32.92 & \underline{0.90} & 29.09 & 0.79 & 27.92 & 0.74 & 27.45 & 0.82 \\
        SwinIR+ & TRAN & 32.93 & \underline{0.90} & 29.15 & 0.79 & 27.95 & 0.75 & 27.56 & 0.83 \\
        SwinFIR & TRAN & 33.20 & $\textbf{0.91}$ & 29.36 & 0.79 & 28.03 & 0.75 & 28.12 & 0.84 \\
        LTE & TRAN & 32.81 &  \texttt{-{}-} & 29.06 &  \texttt{-{}-} & 27.86 &  \texttt{-{}-} & 27.24 &  \texttt{-{}-} \\
        HAT-L & TRAN & \underline{33.30} & \underline{0.90} & 29.47 & \underline{0.80} & 28.09 & \underline{0.76} & 28.60 & \underline{0.85} \\
        DRCT-L & {TRAN} & {\texttt{-{}-}} &  {\texttt{-{}-}} & {\underline{29.54}} &  {\underline{0.80}} & {\underline{28.16}} &  {\underline{0.76}} & {28.70} &  {\underline{0.85}} \\
        {DRCT} & {TRAN} & {\texttt{-{}-}} &  {\texttt{-{}-}} & {29.40} & {\underline{0.80}} & {28.06} &  {0.75} & {28.40} &  {\underline{0.85}} \\
        {HMA} & {TRAN} & {\textbf{33.38}} &  {\textbf{0.91}} & {29.51} & {\underline{0.80}} & {28.13} &  {\underline{0.76}} & {28.69} &  {\underline{0.85}} \\
        {Hi-IR-L} & {TRAN} & {33.22} &  {\textbf{0.91}} & {29.49} & {\underline{0.80}} & {28.13} &  {\underline{0.76}} & {\underline{28.72}} &  {\underline{0.85}} \\
        {CPAT+} & {TRAN} & {33.24} &  {\textbf{0.91}} & {29.36} & {\underline{0.80}} & {28.06} &  {0.75} & {28.33} &  {0.84} \\
        {CPAT} & {TRAN} & {33.19} & {\textbf{ 0.91}} & {29.34} &  {\underline{0.80}} & {28.04} &  {0.75} & {28.22} &  {0.84} \\
        {MaIR} & {TRAN} & {33.14} & {\underline{0.90}} & {29.28} &  {\underline{0.80}} & {\texttt{-{}-}} &  {\texttt{-{}-}} & {27.89} &  {0.83} \\
        {MaIR+} & {TRAN} & {32.93} & {\underline{0.90}} & {29.20} &  {0.79} & {\texttt{-{}-}} &  {\texttt{-{}-}} & {27.71} &  {0.83} \\ 
        {DAT+} & {TRAN} & {33.15} & {\textbf{0.91}} & {29.29} & {\underline{0.80}} & {28.03} & {0.75} & {27.99} & {0.84} \\ 
        {SRFormer+} & {TRAN} & {33.09} & {\textbf{0.91}} & {29.19} & {\underline{0.80}} & {28.00} & {0.75} & {27.85} & {0.84} \\
        \midrule
        SuRGe (Ours) & GAN & 33.07 & \textbf{0.91} & \textbf{30.21} &  \textbf{0.83} & \textbf{31.52} & \textbf{0.87} & \textbf{30.11} & \textbf{0.90} \\
        \bottomrule
        \multicolumn{10}{l}{$^{1}$ Boldfaced: best, Underlined: second best. $^{2}$ CNNA: CNN+Attention, TRAN: Transformer.} \\
        \multicolumn{10}{l}{$^{3}$ Increment indicates improvement.}
    \end{tabular}
\end{table}

\begin{table}[!ht]
    \centering
    \caption{Performance comparison of SuRGe in terms of PSNR and SSIM on six additional test datasets.}
    \label{tab:quant_eval2}
    \scriptsize
    \vspace{5pt}
    \begin{threeparttable}
    \begin{tabular}{llcc}
    \toprule
    Dataset & Method & PSNR($\uparrow$) & SSIM($\uparrow$) \\
    \midrule  
    \multirow{3}{*}{PIRM}
    & ESRGAN+ & 24.15 &  \texttt{-{}-} \\
    & RankSRGAN & \underline{25.62} &  \texttt{-{}-}\\
    &  SuRGe (Ours) & \textbf{31.92} & \textbf{0.90} \\
    \midrule
    \multirow{2}{*}{OST300}
    & ESRGAN+ & \underline{23.80} &  \texttt{-{}-} \\
    & SuRGe (Ours) & \textbf{31.01} & \textbf{0.86} \\
    \midrule
    \multirow{3}{*}{Kitti2012 }
    & NAFSSR-L & \underline{27.12} & \underline{0.82} \\
    & SwinFIR & 26.83 & 0.81 \\
    & SuRGe (Ours) & \textbf{32.31} & \textbf{0.88} \\
    \midrule
    \multirow{3}{*}{Kitti2015 }
    & NAFSSR-L & \underline{26.96} & \underline{0.82} \\
    & SwinFIR & 26.00 & 0.80 \\
    & SuRGe (Ours) & \textbf{31.12} & \textbf{0.89} \\
    \midrule
    \multirow{3}{*}{Middlebury}
    & NAFSSR-L & \underline{30.20} & 0.85 \\
    & SwinFIR & 30.01 & \underline{0.86} \\
    & SuRGe (Ours) & \textbf{35.72} & \textbf{0.93} \\
    \midrule
    \multirow{15}{*}{MANGA109}
    & SRCNN & 27.66 & 0.86 \\
    & DBPN-RES & 31.74 & 0.92 \\
    & HAT-L & 33.09 & \underline{0.93} \\
    & HAT & 32.87 & \underline{0.93} \\
    & SwinFIR & 32.83 & \underline{0.93} \\
    & SwinIR+ & 32.22 & 0.92 \\
    & SAN & 31.66 & 0.92 \\
    & {DRCT} & {32.96} & {\underline{0.93}} \\
    & {DRCT-L} & {33.14} & {\underline{0.93}} \\
    & {HMA} & {\underline{33.19}} & {\underline{0.93}} \\
    & {Hi-IR-L} & {33.13} & {\underline{0.93}} \\
    & {CPAT+} & {32.85} & {\underline{0.93}} \\
    & {MAIR} & {32.46} & {\underline{0.93}} \\
    & {MAIR+} & {32.66} & {\underline{0.93}} \\
    & SuRGe (Ours) & \textbf{34.17} & \textbf{0.95} \\
    \bottomrule
    \end{tabular}
    \begin{tablenotes}
        \item The best result is boldfaced, while the second best is underlined.
    \end{tablenotes}
    \end{threeparttable}
\end{table}

 \subsection{Quantitative performance of SuRGe} In Table \ref{tab:quant_eval} we exhibit the efficacy of the proposed SuRGe on Set5, Set14, BSD100, and Urban100 benchmarks in terms of PSNR and SSIM. For comparison, we select 26 state-of-the-art methods from 4 groups. \textbf{(1)} CNN-based: SRCNN \cite{dong2015srcnn}, SelfExSR \cite{urban100}, and DBPN-RES-MR64-3 \cite{dbpntpami2021}. \textbf{(2)} GAN-based: SRGAN \cite{srgan_cvpr2017}, ProSR-L \cite{wang2018prosrl}, ESRGAN \cite{wang2018esrgan}, RankSRGAN \cite{Zhang_2019_ranksrgan}, Beby-GAN \cite{li2022bybygan}, and GramGAN \cite{gramgansong_2023}. \textbf{(3)} CNN with attention-based: SAN \cite{sancvpr2019} and WRAN \cite{wran_neurocomputing2020}. \textbf{(4)} Transformer-based: SwinIR and SWIN-IR+ \cite{swiniriccvw2021}, LTE \cite{lee2022lte}, SwinFIR \cite{zhang2022swinfir}, HAT-L \cite{chen2205hat-l}, DRCT and DRCT-L \cite{hsu2024drct}, HMA \cite{chu2024hmanet}, Hi-IR-L \cite{li2024hiirl}, CPAT and CPAT+ \cite{tran2024cpat}, MaIR and MaIR+ \cite{li2024mair}, DAT+ \cite{daticcv2023}, and SRFormer+ \cite{srformericcv2023}. Table \ref{tab:quant_eval} shows that, except for Set5, on all the other datasets, SuRGe achieves a higher PSNR and SSIM than the contending methods, attesting to its highly consistent better performance. Specifically, on average, on the three datasets, namely, Set14, BSD100, and Urban100, the proposed SuRGe improves the PSNR and SSIM metrics over the corresponding state-of-the-art by a respective margin of 6.28\% and 7.5\%. In the case of Set5, SuRGe achieves the best SSIM jointly with five recent transformer-based methods, namely SwinFIR, HMA, Hi-IR-L, CPAT, and CPAT+. When the PSNR is considered, SuRGe stands at the ninth position among the 27 contenders on Set5, lagging closely behind eight transformer-based techniques, although the actual PSNR achieved by SuRGe is only 0.93\% lower than the corresponding state-of-the-art method HMA. In other words, specifically on Set5, the proposed SuRGe maintains a closely comparable performance with the recent transformer-based techniques, if not being able to outperform them. This peculiarity on Set5 may be attributed to the greater difficulty inherent in the dataset due to its exceptionally small LR images. The LR images in Set5 are not only considerably smaller than the DIV2K training and other testing datasets used in this study, but also often contain intricate details. In other words, densely packed information in extremely low-resolution makes them harder to correctly reproduce (and enhance) in the SR for Set5, especially through a generative network that is not purposefully trained on such examples. Consequently, the SR output of SuRGe retains most of the high-level visual similarity to attain a commendable SSIM, while the loss of some minute details is apparent from the slightly lower PSNR. On the other hand, one cannot surely conclude that the transformer-based methods, by virtue of the attention block, are more inherently suited for such anomalous usage; given among the 13 such contenders, five achieve a lower PSNR, and eight attain a lower SSIM than the proposed GAN-based SuRGe.

 We further evaluate the efficacy of SuRGe on Kitti2012, Kitti2015, Middlebury, PIRM, OST300, and MANGA109, a set of modern super-resolution benchmarks that focus on a high level of detailing, color, and contrast. We compare the performance of SuRGe in terms of PSNR and SSIM in Table \ref{tab:quant_eval2} against 17 notable contenders, viz. SRCNN, DBPN-RES-MR64-3, ESRGAN+ \cite{rakotonirina2020esrgan+}, SAN, RankSRGAN \cite{Zhang_2019_ranksrgan}, NAFSSR-L \cite{chu2022nafssr}, SwinIR+, SwinFIR, HAT and HAT-L \cite{chen2205hat-l}, DRCT and DRCT-L \cite{hsu2024drct}, HMA \cite{chu2024hmanet}, Hi-IR-L \cite{li2024hiirl}, CPAT+ \cite{tran2024cpat}, MAIR and MAIR+ \cite{li2024mair}. We see from Table \ref{tab:quant_eval2} that SuRGe achieves better PSNR and SSIM on all six datasets. Specifically, for PSNR, the proposed SuRGe shows a commendable 17.59\% improvement on average. This establishes the power of SuRGe in consistently generating better quality SR outputs.

 \subsection{Qualitative comparison} 
\begin{figure}[!ht]
	\centering
		\includegraphics[width=0.8\textwidth]{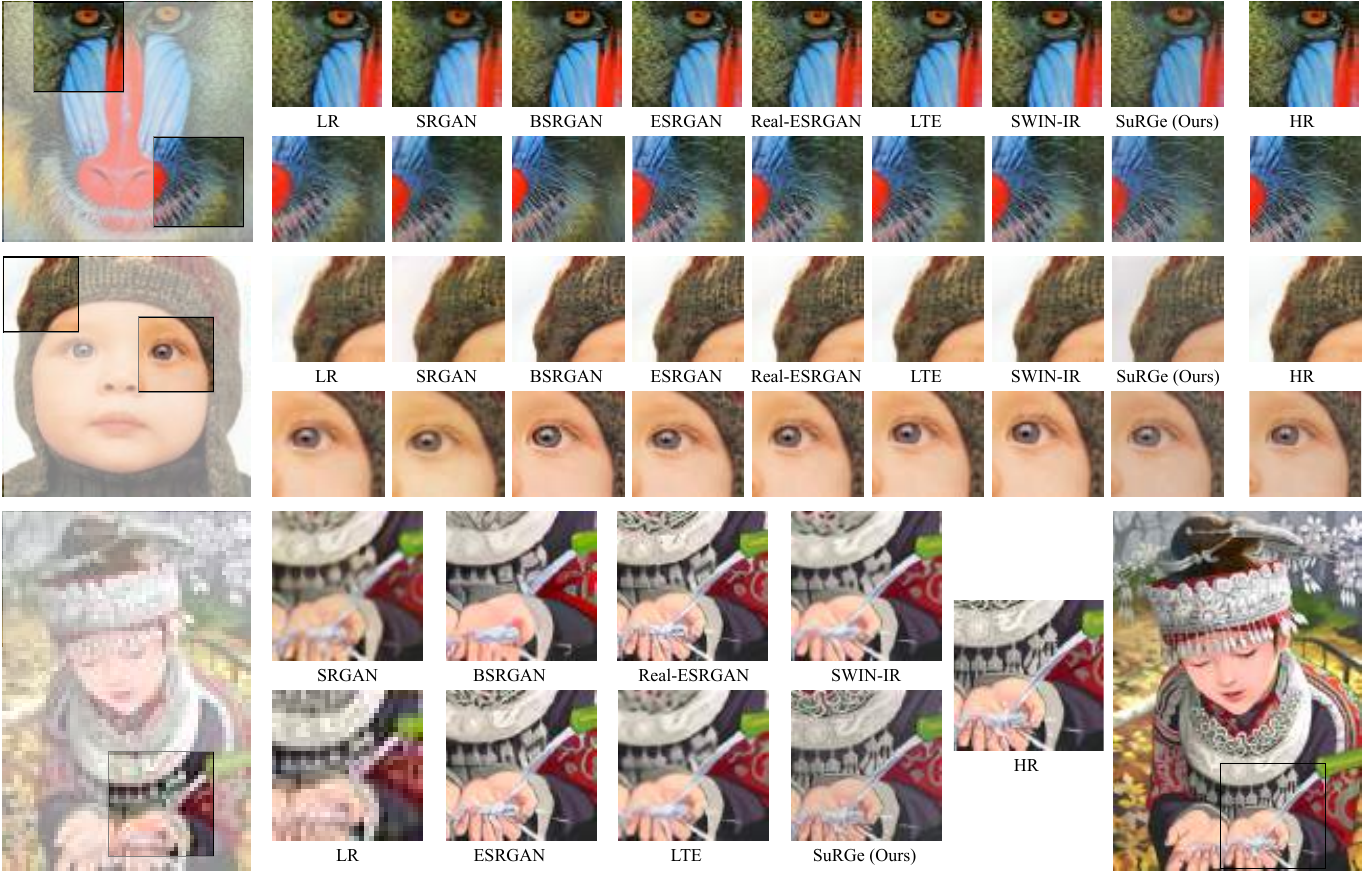}
		\caption{A qualitative comparison of SR outputs generated by SRGAN \cite{srgan_cvpr2017}, BSRGAN \cite{gu2019bsrgan}, ESRGAN \cite{wang2018esrgan}, Real-ESRGAN \cite{wang2021realesrgan}, LTE \cite{lee2022lte}, SWIN-IR \cite{swiniriccvw2021} and SuRGe (Ours) in baboon (Set14), child (Set5) and comic (Set14) samples. The SR output of SuRGe is visually more similar to the HR ground truth for almost every patch for the three images.}
  \label{fig:visual_comp}
\end{figure}

We compare the visual quality of SuRGe against BSRGAN \cite{gu2019bsrgan}, SRGAN, ESRGAN \cite{wang2018esrgan}, Real-ESRGAN \cite{wang2021realesrgan}, LTE, and SwinIR in Figure \ref{fig:visual_comp}. From Figure \ref{fig:visual_comp}, we can make three key observations. \textbf{(1)} SRGAN, LTE, and SWIN-IR output comparatively blurry SR than SuRGe. \textbf{(2)} ESRGAN and Real-ESRGAN, though they preserve finer details, may add more distortion and noise to SR. This is apparent from the mustache of the baboon, the eyebrow of the child, and the nails of the comic. \textbf{(3)} BSRGAN provides smooth, apparently attractive SR outputs but fails to conserve details to the limit of SuRGe. Thus, SuRGe produces better and more detailed SR outputs closer to the HR ground truths. Additional qualitative results along with quantitative support in favor of SuRGe can be found in \ref{app:sec:additionalResults}.

\section{\textbf{Conclusion and Future Works}}
In this study, we propose a fully-convolutional GAN-based method called SuRGe that generates visually attractive $4$x super-resolution images with minute details. In essence, SuRGe stresses the need for diverse informative feature preservation and their combination in a learnable fashion in the super-resolution task. Further, possibly for the first time, SuRGe considers the importance of aligning the distributions of SR and HR by successfully applying a suitable divergence measure, such as GW, as a loss function in the super-resolution context. Moreover, SuRGe demonstrates how the key architectural choices, such as kernel size, normalization methods, and the location and strategy of up-scaling, altogether impact the quality of the generated SR output. Such judicious use of intuition-driven heuristics allows SuRGe to achieve a commendable performance with a smaller model and consequently a lower inference time in comparison to typical GAN-based SR methods (see Section \ref{app:sec:additionalResults} in Appendix for details). 
 
One limitation of SuRGe stems from its heavy reliance on divergence measures. Specifically, divergence measures are often susceptible to noise present in the SR, which may potentially compromise the quality of the SuRGe-generated SR. An interesting future direction of research may attempt to empirically evaluate and improve the robustness of SuRGe by either exploring the applicability of robust divergence measures \cite{he2003generalized} or incorporating additional remedies like the median of means \cite{GuillaumeMom2020}. Moreover, the currently proposed architecture is tailored for 4x super-resolution, as that is the most common and widely popular variant of the task. Thus, a direct extension of SuRGe may focus on generalizing to an $r$x super-resolution task. In this case, one may investigate the corrective measures required to seamlessly preserve and propagate intricate information through repetitive up-sampling stages while maintaining scalability. An allied avenue may look for an up-sampling method that is more suitable for super-resolution, e.g., that induces minimal distortion or performs self-correction to mitigate information loss. Further, the recent diffusion models are slowly but surely gaining popularity for the super-resolution task. Therefore, an interesting research may seek to apply the learning from SuRGe, especially the use of divergence measures, in diffusion-based models \cite{wang2024sinsr}. Such an extension is also possible for transformers, where curated attention schemes may strongly couple with the efficient loss functions.

\newpage
\appendix

\section{Detailed architecture of the SuRGe network}
\label{app:sec:networkDetails}
\begin{figure}[!ht]
	\centering
		\includegraphics[width=0.9\textwidth]{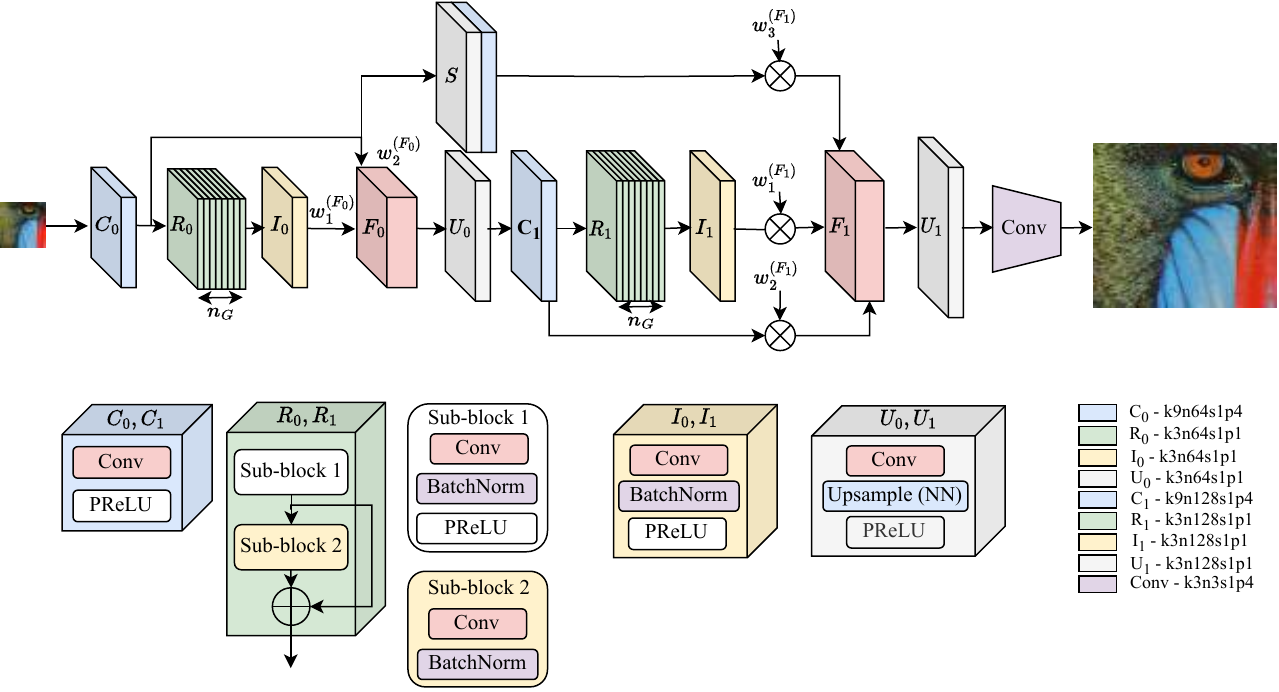}
            \caption{The architecture of $G$ in SuRGe with the different individual components described in detail.}
  \label{fig:generator}
\end{figure}

Refer to Figure \ref{fig:generator} and \ref{fig:discriminator} for the detailed schematic description of the respective architectures of Generator $G$ and Discriminator $D$ in SuRGe.

\subsection{Architecture of Generator}
We present the architecture of Generator $G$ in SuRGe. Taking a random LR patch from Baboon as an example input, the model constructed with $n_G = 8$ outputs a SR patch by passing it through different blocks as indicated by the legend.

\subsection{Architecture of Discriminator}
\begin{figure}[!ht]
	\centering
		\includegraphics[width=.8\textwidth]{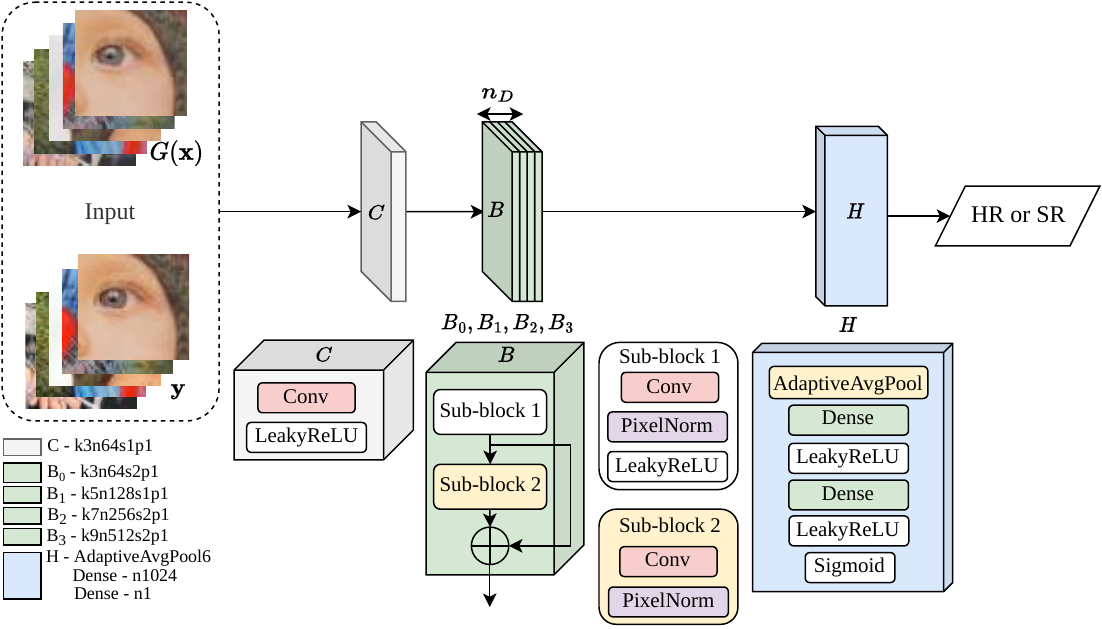}	
            \caption{The architecture of $D$ in SuRGe with a detailed description of the individual components is presented.}
  \label{fig:discriminator}
\end{figure}

We demonstrate the discriminator model $D$ of SuRGe in Figure \ref{fig:discriminator}. All blocks presented in the figure follow the same naming convention discussed in Section \ref{sec:discriminator} of the main paper. Discriminator $D$ uses $n_D = 4$ along with an architecture of sub-blocks that is structurally similar to the generator $G$, except for the presence of normalisation. The classification head $H$ is responsible for distinguishing between real (HR) and fake (SR) image samples.

\section{Algorithm of SuRGe}
\label{app:sec:algo}
The following Algorithm \ref{alg:surge} describes the workflow of SuRGe.
\begin{algorithm}[!ht]
  \caption{\footnotesize Super-Resolution Generator (SuRGe)}
  \label{alg:surge}
  \scriptsize
  {\bfseries Input:} $Y^{GT}=\{\mathbf{y}^{GT}_{1}, \mathbf{y}^{GT}_{2}, \cdots, \mathbf{y}^{GT}_{m}\}$: training set of full HR Ground Truth (GT) images, $N$: mini-batch size, $T$: Number of epochs as a termination criterion. \\
  {\bfseries Output:} A trained super-resolution image generator network $G$. 
  \let \oldnoalign \noalign
  \let \noalign \relax
  \midrule
  \let \noalign \oldnoalign
  \begin{algorithmic}[1]
  \STATE Initialize epoch counter $t = 1$.
  \WHILE{$t \leq T$} 
    \STATE Initialize $Y=\phi$.
    \FOR{Each of $\mathbf{y}^{GT} \in Y^{GT}$}
        \STATE $Y = Y \cup \{\mathbf{y}\}$, where $\mathbf{y}$ is a $256 \times 256$ patch, randomly extracted from $\mathbf{y}^{GT}$. 
    \ENDFOR
    \STATE Sample $Y_{N} = \{\mathbf{y}_{1}, \mathbf{y}_{2}, \cdots, \mathbf{y}_{N}\} \subset Y$ of a batch of HR ground truth patches.
    \STATE Form $X_{N} = \{\mathbf{x}_{1}, \mathbf{x}_{2}, \cdots, \mathbf{x}_{N}\}$ of LR training batch where $\mathbf{x}_{i}$ is formed by down-scaling $\mathbf{y}_{i}$ to $64 \times 64$ by bi-cubic interpolation for all $i=1, 2, \cdots. N$.
    \STATE Update $G$ by gradient descent on $\mathcal{L}^{G}(X_{N}, Y_{N}, G(X_{N}))$.
    \STATE Sample SR output batch $G(X_{N})= \{G(\mathbf{x}_{1}), G(\mathbf{x}_{2}), \cdots, G(\mathbf{x}_{N})\}$ where $\mathbf{x}_{i} \in X_{n}$.
    \STATE Update $D$ by gradient descent on $\mathcal{L}^{D}(Y_{N}, G(X_{N}))$.
    \STATE Increase epoch counter $t = t+1$.
   \ENDWHILE
\end{algorithmic}
\end{algorithm}

\section{Description of datasets}
\label{app:sec:dataDescription}
The following Table \ref{tab:data_stats} provides the details of the DIV2K training dataset \cite{DIV2K} and the 10 benchmark testing datasets. 
\begin{table}[!ht]
    \scriptsize
    \centering
    \caption{Details of datasets}
    \label{tab:data_stats}
    \vspace{5pt}
    \begin{tabular}{l|cc|p{6cm}} \toprule
       Dataset & Number of & Average ground & Remark \\
       & samples & truth resolution & \\ \midrule
       DIV2K \cite{DIV2K} & $800$ & $1971 \times 1435$ & Only the training split is used to train SuRGe. \\ \midrule
        Set5 \cite{set5} & $5$ & $78 \times 84$ & Benchmark, contains 5 samples in total. \\
        Set14 \cite{set14} & $14$ & $112 \times 101$ & Benchmark, contains 14 samples in total. \\
        BSD100 \cite{bsd100} & $100$ & $110 \times 89$ & Benchmark, contains 100 samples in total. \\
        Urban100 \cite{urban100} & $100$ & $246 \times 199$ & Benchmark, contains 100 samples in total. \\
        PIRM \cite{pirm} & $100$ & $155 \times 119$ & contains 100 samples from validation set. \\
        KITTI2012 \cite{kittidataset} & $20$ & $300 \times 96$ & The same 20 samples are taken as in \cite{chu2022nafssr}. \\
        KITTI2015 \cite{kittidataset} & $20$ & $300 \times 96$ & The same 20 samples are taken as in \cite{chu2022nafssr}. \\
        Middlebury \cite{middlebury} & $10$ & $438 \times 312$ & The same 10 samples are taken as in \cite{chu2022nafssr}. \\
        OST300 \cite{wang2018ost300} & $300$ & $160 \times 123$ & contains 300 samples in total. \\
        MANGA109 \cite{manga109} & $109$ & $207 \times 300$ & contains 109 samples in total. \\
        \bottomrule
    \end{tabular}
\end{table}

\section{Metrics}
\label{app:sec:metrics}
We use two metrics to quantify and compare the performance of the super-resolution methods.
\subsection{Peak Signal to Noise Ratio (PSNR)}
PSNR \cite{PSNR} is a measure to quantify the noise present in an image compared to a reference. In case of super-resolution, if the ground truth is $\mathbf{y}$ and the SR output generated by $G$ from a LR input $\mathbf{x}$ is $G(\mathbf{x})$, then the PSNR ($\rho$) between the two is defined as:
\begin{equation}
    \label{eq:psnr}
    \rho(\mathbf{y}, G(\mathbf{x})) = 20 \log_{10}\left[\frac{\max \{\mathbf{y}\}}{\frac{1}{wh}||\mathbf{y}-G(\mathbf{x})||^{2}_{2}}\right],
\end{equation}
where $w$ and $h$ respectively denote the width and height or $\mathbf{y}$ or $G(\mathbf{x})$. Evidently, in equation (\ref{eq:psnr}), we want to decrease $||\mathbf{y}-G(\mathbf{x})||^{2}_{2}$ so that the SR output matches the HR ground truth. In other words, a higher PSNR indicates a better quality SR. 

\subsection{Structural Similarity Index (SSIM)}
SSIM \cite{wang2004ssim} is another commonly used metric to compare an SR output to a reference HR ground truth. Similar to PSNR, given a SR output $G(\mathbf{x})$ and a HR ground truth $\mathbf{y}$ the SSIM $\lambda$ between the two is calculated as follows:
\begin{equation}
\lambda(\mathbf{y}, G(\mathbf{x})) = \frac{\left(2\mu_{\mathbf{y}}\mu_{G(\mathbf{x})} + c_1\right)\left(2Cov(\mathbf{y}, G(\mathbf{x})) + c_2\right)}{\left(\mu_\mathbf{y}^2 + \mu^2_{G\left(\mathbf{x}\right)} + c_1\right)\left(\sigma_\mathbf{y}^2 + \sigma^2_{G\left(\mathbf{x}\right)} + c_2\right)},
\end{equation}
where, $\mu_\mathbf{y}$ and $\mu_{G(\mathbf{x})}$ are respectively the mean pixel values of HR ground truth and SR output, $\sigma_\mathbf{y}^2$ and $\sigma^2_{G\left(\mathbf{x}\right)}$ denote the variance of pixel values in $\mathbf{y}$ and $G(\mathbf{x})$, while $Cov(\mathbf{y}, G(\mathbf{x}))$ is the covariance between the pixel values of the two images. The two constants $c_{1}$ and $c_{2}$ ensure the non-zero property for the denominator and are usually kept to small values. A higher SSIM indicates the two images are more perceptually similar. 

\section{Network architecture selection by grid search}
\label{app:sec:networkSearch}
The grid search spaces and the final selected networks, along with hyper-parameter settings for the generator $G$ and the discriminator $D$ in SuRGe are respectively detailed in Tables \ref{tab:surgeG} and \ref{tab:surgeD}.

\begin{table}[!ht]
    \centering
    \scriptsize
    \vspace{5pt}
    \caption{Grid search space with the selected $G$ architecture and hyperparameter settings of SuRGe.}
    \label{tab:surgeG}
    \begin{threeparttable}
    \begin{tabular}{llll}
    \toprule
    Block & Parameters & Grid search space & Final network \\
    \midrule
    \multirow{8}{*}{$C_{0}, C_{1}$} & Kernel size ($C_{0}, C_{1}$) & $\{7, 9, 11\}$ & 9 \\
    & No. of convolutional filters ($C_{0}$) & $\{32, 64\}$ & 64\\
    & No. of convolutional filters ($C_{1}$) & $\{64, 128\}$ & 128\\
    & Stride ($C_{0}, C_{1}$) & 1 & 1\\
    & Padding ($C_{0}, C_{1}$) & $\{1, 2, 4\}$ & 4 \\
    & Activation ($C_{0}, C_{1}$) & ReLU, LeakyReLU, PReLU & PReLU \\
    & Normalization ($C_{0}, C_{1}$) & True, False & False \\
    & Normalization technique ($C_{0}, C_{1}$) & \texttt{-{}-} & \texttt{-{}-} \\
    & No. of convolution layers ($C_{0}, C_{1}$) & $\{1, 2\}$ & 1 \\ \midrule
    \multirow{10}{*}{$R_{0}, R_{1}$} & Inter-sub-blocks skip connection ($R_{0}, R_{1}$) & True, False & True \\
    & No. of convolutions in sub-block ($R_{0}, R_{1}$) & $\{1, 2\}$ & 1 \\
    & No. of convolution filters ($R_{0}$) & $\{32, 64\}$ & 64 \\
    & No. of convolution filters ($R_{1}$) & $\{64, 128\}$ & 128 \\
    & Kernel size ($R_{0}, R_{1}$) & $\{3, 5\}$ & 3 \\
    & (Stride, Padding) ($R_{0}, R_{1}$) & (1, 1) & (1, 1) \\
    & Activation presence in sub-blocks ($R_{0}, R_{1}$) & First, Second, Both & First \\
    & Normalization in sub-blocks ($R_{0}, R_{1}$) & First, Second, Both & Both \\
    & Activation technique ($R_{0}, R_{1}$) & ReLU, LeakyReLU, PReLU & PReLU \\
    & Normalization technique ($R_{0}, R_{1}$) & BatchNorm, PixelNorm & BatchNorm \\ 
    \midrule
    \multirow{8}{*}{$I_{0}, I_{1}$} & Kernel size ($I_{0}, I_{1}$) & $\{3, 5\}$ & 3 \\
    & No. of convolutional filters ($I_{0}$) & $\{32, 64\}$ & 64 \\
    & No. of convolutional filters ($I_{1}$) & $\{64, 128\}$ & 128 \\
    & (Stride, Padding) ($I_{0}, I_{1}$) & (1, 1) & (1, 1) \\
    & Activation ($I_{0}, I_{1}$) & ReLU, LeakyReLU, PReLU & PReLU \\
    & Normalization ($I_{0}, I_{1}$) & True, False & True \\
    & Normalization technique ($I_{0}, I_{1}$) & BatchNorm, PixelNorm & BatchNorm \\
    & No. of convolution layers ($I_{0}, I_{1}$) & $\{1, 2\}$ & 1 \\ \midrule
    \multirow{10}{*}{$U_{0}, U_{1}$} & Initial convolution present & True, False & True \\
    & Kernel size of convolution ($U_{0}, U_{1}$) & $\{3, 5\}$ & 3 \\
    & No. of convolutional filters ($U_{0}$) & $\{32, 64\}$ & 64 \\
    & No. of convolutional filters ($U_{1}$) & $\{64, 128\}$ & 128 \\
    & (Stride, Padding) ($U_{0}, U_{1}$) & (1, 1) & (1, 1) \\
    & Activation ($U_{0}, U_{1}$) & ReLU, LeakyReLU, PReLU & PReLU \\
    & Normalization ($U_{0}, U_{1}$) & True, False & False \\
    & Normalization technique ($U_{0}, U_{1}$) & \texttt{-{}-} & \texttt{-{}-} \\
    & No. of convolution layers ($U_{0}, U_{1}$) & $\{1, 2\}$ & 1 \\ 
    & Up-scaling technique ($U_{0}, U_{1}$) & Bi-cubic, PixelShuffle, NN$^{*}$ & NN \\ \midrule
    $S$ & Structure similar to & $U_{0} \circ C_{1}$ & $U_{0} \circ C_{1}$ \\ \midrule
    Optimizer & Learning rate & $\{0.0001, 0.00025, 0.0005\}$ & 0.0001 \\
    Adam & ($\beta_{1}, \beta_{2}$) & (0.9, 0.99) & (0.9, 0.99) \\ 
    \bottomrule
    \end{tabular}
    \begin{tablenotes}
    \item NN$^{*}$: Nearest Neighbour method.
    \end{tablenotes}
    \end{threeparttable}
\end{table}

\begin{table}[!ht]
    \centering
    \scriptsize
    \caption{Grid search space with the selected $D$ architecture and hyperparameter settings of SuRGe.}
    \vspace{5pt}
    \label{tab:surgeD}
    \begin{threeparttable}
    \begin{tabular}{llll}
    \toprule
    Block & Parameters & Grid search space & Final network \\
    \midrule
    \multirow{8}{*}{$C$} & Kernel size & $\{3, 5\}$ & 3 \\
    & No. of convolutional filters & $\{32, 64\}$ & 64\\
    & (Stride, Padding) & (1, 1) & (1, 1) \\
    & Activation & ReLU, LeakyReLU, PReLU & LeakyReLU \\
    & Leakiness of LeakyReLU & $\{0.1, 0.2\}$ & 0.2 \\
    & Normalization & True, False & False \\
    & Normalization technique & \texttt{-{}-} & \texttt{-{}-} \\
    & No. of convolution layers & $\{1, 2\}$ & 1 \\ \midrule
    \multirow{11}{*}{$B$} & Inter-sub-blocks skip ($B$) & True, False & True \\
    & No. of $B$ blocks & $\{1, 2, 3, 4\}$ & 4 \\
    & No. of conv in sub ($B_{0}-B_{3}$) & $\{1, 2\}$ & 1 \\
    & No. of conv filters ($B_{0}-B_{3}$) & $(64, 128, 256, 512)$ & $(64, 128, 256, 512)$ \\
    & Kernel size ($B_{0}-B_{3}$) & (3, 5, 7, 9) & (3, 5, 7, 9) \\
    & (Stride, Padding) ($B$) & (1, 1) & (1, 1) \\
    & Activation in sub ($B$) & First, Second, Both & First \\
    & Normalization in sub ($B$) & First, Second, Both & Both \\
    & Activation technique ($B$) & ReLU, LeakyReLU, PReLU & LeakyReLU \\
    & Leakiness of LeakyReLU ($B$) & $\{0.1, 0.2\}$ & 0.2 \\
    & Normalization technique ($B$) & PixelNorm & PixelNorm \\ 
    \midrule
    \multirow{8}{*}{$H$} & Pooling strategy & avgPool, adaptiveAvgPool, maxPool & adaptiveAvgPool \\
    & Output size of adaptiveAvgPool & $\{4, 6, 8\}$ & 6 \\
    & No. of dense layers & $\{1, 2, 3\}$ & 2 \\
    & No. of dense nodes & $\{(512, 1), (1024, 1), (2048, 1)\}$ & (1024, 1) \\
    & Activation & ReLU, LeakyReLU, PReLU & LeakyReLU \\
    & Leakiness of LeakyReLU & $\{0.1, 0.2\}$ & 0.2 \\
    & Normalization & True, False & True \\
    & Normalization technique & \texttt{-{}-} & \texttt{-{}-} \\ \midrule
    Gradient & Weight $\gamma$ & $\{1, 10, 100\}$ & 10 \\
    Penalty & & & \\ \midrule
    Optimizer & Learning rate & $\{0.0001, 0.00025, 0.0005\}$ & 0.0001 \\
    Adam & ($\beta_{1}, \beta_{2}$) & (0, 0.9) & (0, 0.9) \\ 
    \bottomrule
    \end{tabular}
    \end{threeparttable}
\end{table}

\section{Additional Results}
\label{app:sec:additionalResults}
\subsection{Inference time of SuRGe compared to recent GAN and Transformer-based contenders} 
To confirm if the several improvements made in SuRGe maintain a low inference time, we compare the same in seconds against that of three recent GAN and transformer models, viz. BSRGAN, SwinFIR, and LTE in the same computing setup. In essence, all algorithms are executed on a computing system with a single AMD Ryzen 9 3900x 12-core processor, one NVIDIA RTX 3090 24GB GPU, and a total of 64GB DDR4 memory. The following Table \ref{tab:inferenceTime} shows that, on average, on three datasets, namely Set5, Set14, and BSD100, the proposed SuRGe provides good quality SR at the lowest inference time. 

\begin{table}[!ht]
    \centering
    \caption{Average inference time in seconds for super-resolution of a single image. The best result is boldfaced, and the second best is underlined.}
    \label{tab:inferenceTime}
    \vspace{5pt}
    \scriptsize
    \begin{tabular}{l|ccc} \toprule
        Model & Set5 & Set14 & BSD100 \\ \midrule
        BSRGAN & \underline{0.059} & \underline{0.095} & \underline{0.077} \\
        LTE & 0.080 & 0.132 & 0.099 \\
        SwinFIR & 1.032 & 1.886 & 1.389 \\ 
        \midrule
        SuRGe (Ours) & \textbf{0.055} & \textbf{0.089} & \textbf{0.067} \\ \midrule
        Speed-up of SuRGe from the current best model & \textcolor{green}{1.07x} & \textcolor{green}{1.07x} & \textcolor{green}{1.15x} \\
        \bottomrule
    \end{tabular}
\end{table}

\subsection{A comparison of performance vs. number of parameters for SuRGe and its notable GAN-based contenders}
In the following Table \ref{tab:parameters} we provide a comparative study of performance (averaged over the four benchmark datasets viz. Set5 \cite{set5}, Set14 \cite{set14}, BSD100 \cite{bsd100}, and URBAN100 \cite{urban100}, in terms of PSNR and SSIM) vs. the number of parameters (in millions) for GAN-based super-resolution techniques like the proposed SuRGe, PROSR-L \cite{wang2018prosrl}, SRGAN \cite{srgan_cvpr2017}, ESRGAN \cite{wang2018esrgan}, BeByGAN \cite{li2022bybygan}, Rank-SRGAN \cite{Zhang_2019_ranksrgan}. We see from Table \ref{tab:parameters} that even though SuRGe uses about $2\times$ parameters of PROSR-L, it improves the PSNR with about 2.55 percentage points (pp) on average. Moreover, SuRGe and SRGAN use almost the same number of parameters, while the proposed outperforms the contender by about 4pp in PSNR and 13pp in SSIM. The rest of the competing methods use about 1.5x-2x parameters of SuRGe while achieving a lower PSNR and SSIM on average. Thus, SuRGe can be considered a comparatively lightweight GAN-based method that demonstrates high performance while maintaining a limit on the number of parameters. 

\begin{table}[!ht]
    \centering
    \scriptsize
    \caption{Comparison of performance vs. the number of parameters for GAN-based approaches.}
    \label{tab:parameters}
    \vspace{5pt}
    \begin{threeparttable}
    \begin{tabular}{l|c|cc}
    \toprule
    GAN-based Technique & No. of Parameters & \multicolumn{2}{c}{Average Performance} \\
    \cmidrule{3-4}
    & (in millions) & PSNR($\uparrow$) & SSIM($\uparrow$)\\
    \midrule
    PROSR-L \cite{wang2018prosrl} & 13 (\textcolor{green}{$\approx0.5\times$})	& 27.78	(\textcolor{red}{$-4.33$}) & \texttt{-{}-} \\
    SRGAN \cite{srgan_cvpr2017}	& 25.1 (\textcolor{red}{$\approx1\times$}) & 26.19 (\textcolor{red}{$-5.92$}) & 0.75 (\textcolor{red}{$-0.14$}) \\
    ESRGAN \cite{wang2018esrgan} & 40 (\textcolor{red}{$\approx1.5\times$}) & 29.15	(\textcolor{red}{$-2.96$}) & 0.81 (\textcolor{red}{$-0.08$}) \\
    BeByGAN	\cite{li2022bybygan} & 40 (\textcolor{red}{$\approx1.5\times$}) & 26.57	(\textcolor{red}{$-5.54$}) & 0.74 (\textcolor{red}{$-0.15$}) \\
    Rank-SRGAN \cite{Zhang_2019_ranksrgan}	& 53 (\textcolor{red}{$\approx2\times$}) & 26.07 (\textcolor{red}{$-6.04$}) & 0.65 (\textcolor{red}{$-0.24$}) \\ \midrule
    SuRGe (Ours)	& \textbf{25.7} & \textbf{32.11}	& \textbf{0.89} \\
    \bottomrule
    \end{tabular}
    \begin{tablenotes}
        \item \textcolor{green}{Green} indicates the number of parameters/performance of the contender is better than SuRGe. \textcolor{red}{Red} indicates the number of parameters/performance of the contender is worse than SuRGe.
        \item The ratio of the number of parameters and the difference in performance is measured considering SuRGe as a reference. 
    \end{tablenotes}
    \end{threeparttable}
\end{table}

\subsection{Simultaneous demonstration of qualitative and quantitative performances of SuRGe}
\begin{figure}[!h]
	\centering
		\includegraphics[width=0.98\linewidth]{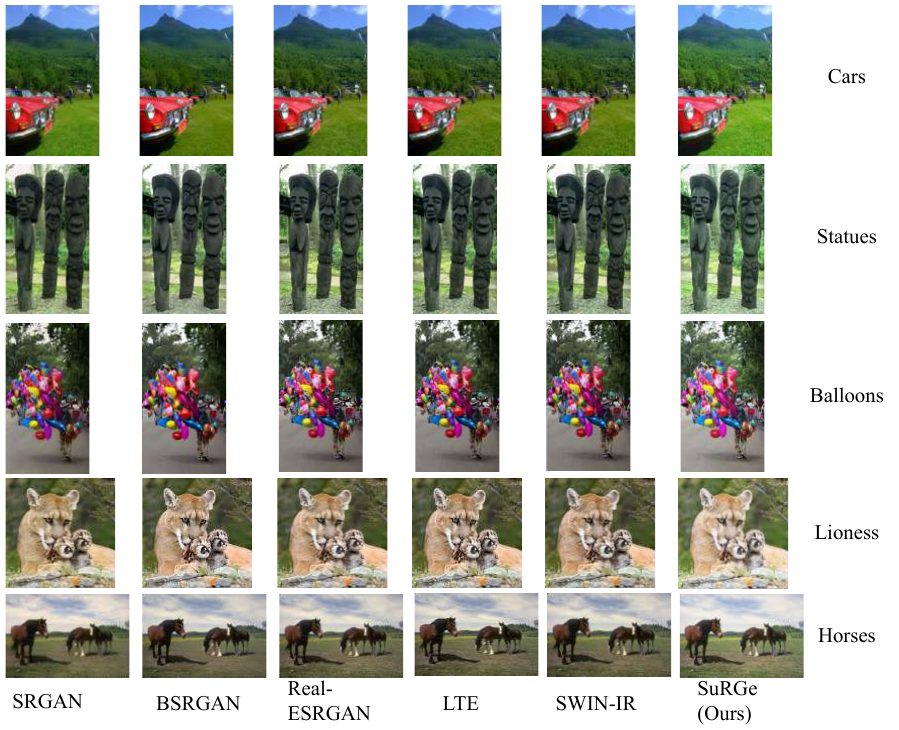}
            \caption{Additional qualitative performance comparison of SuRGe. The proposed SuRGe consistently produces SR outputs with finer details.}
            \label{fig:qual}
\end{figure}

\begin{figure*}[!ht]
    \centering
    \includegraphics[width=\textwidth]{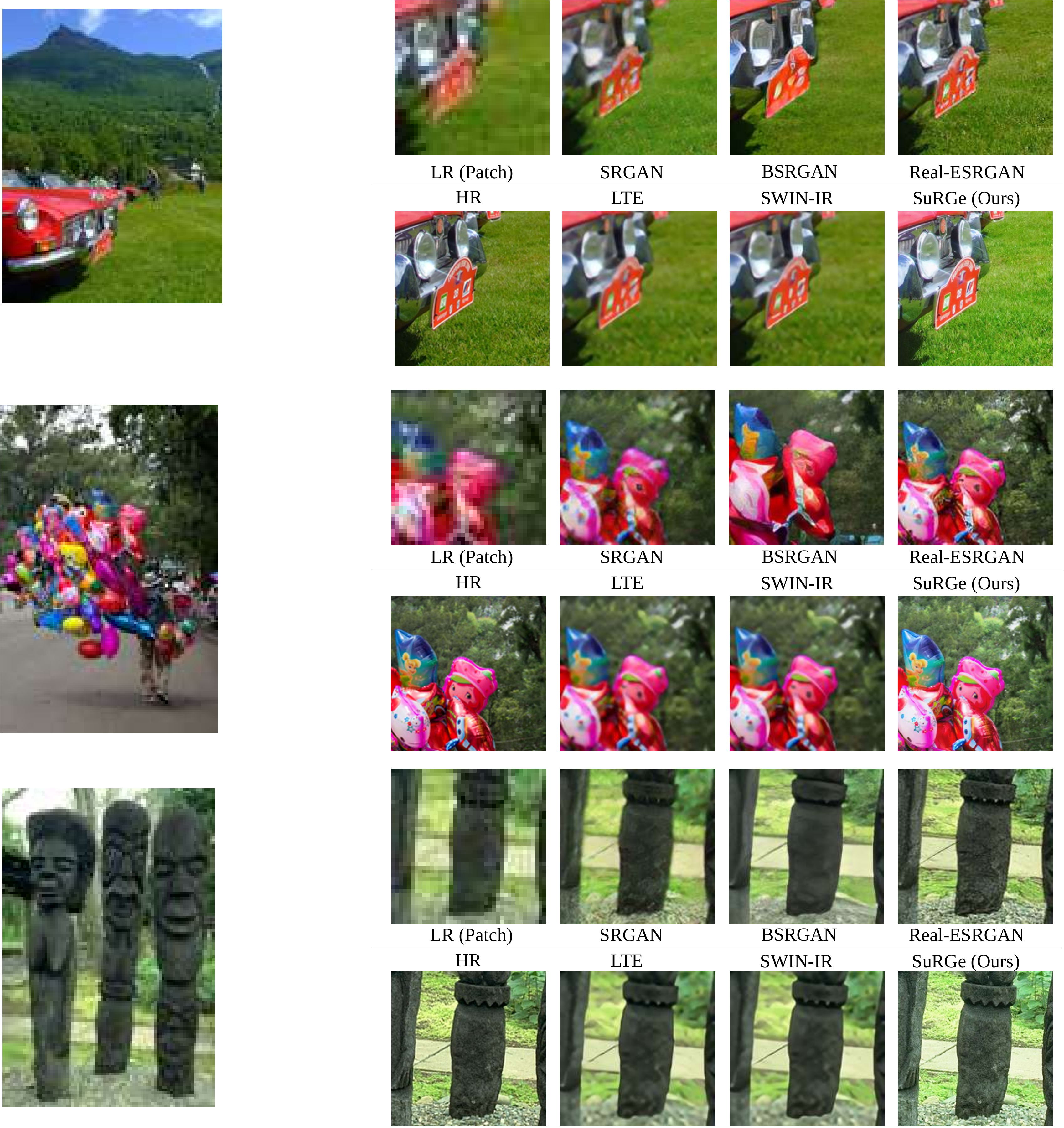}
    \caption{We take three test samples, namely Cars (PIRM), Balloons (PIRM), and Statues (BSD100). We compare SuRGe with the contenders like SRGAN \cite{srgan_cvpr2017}, BSRGAN \cite{gu2019bsrgan}, Real-ESRGAN \cite{wang2021realesrgan}, LTE \cite{lee2022lte}, and SWIN-IR \cite{swiniriccvw2021}. A patch-based visual comparison further aids us in clearly observing the greater amount of minute details (sunlight reflection on the car headlight, eye details of the balloon toys, stone texture in the statue) preserved by SuRGe in the SR output. }
    \label{fig:addtional_qual}
\end{figure*}

\begin{figure*}[!ht]
    \centering
    \includegraphics[width=\textwidth]{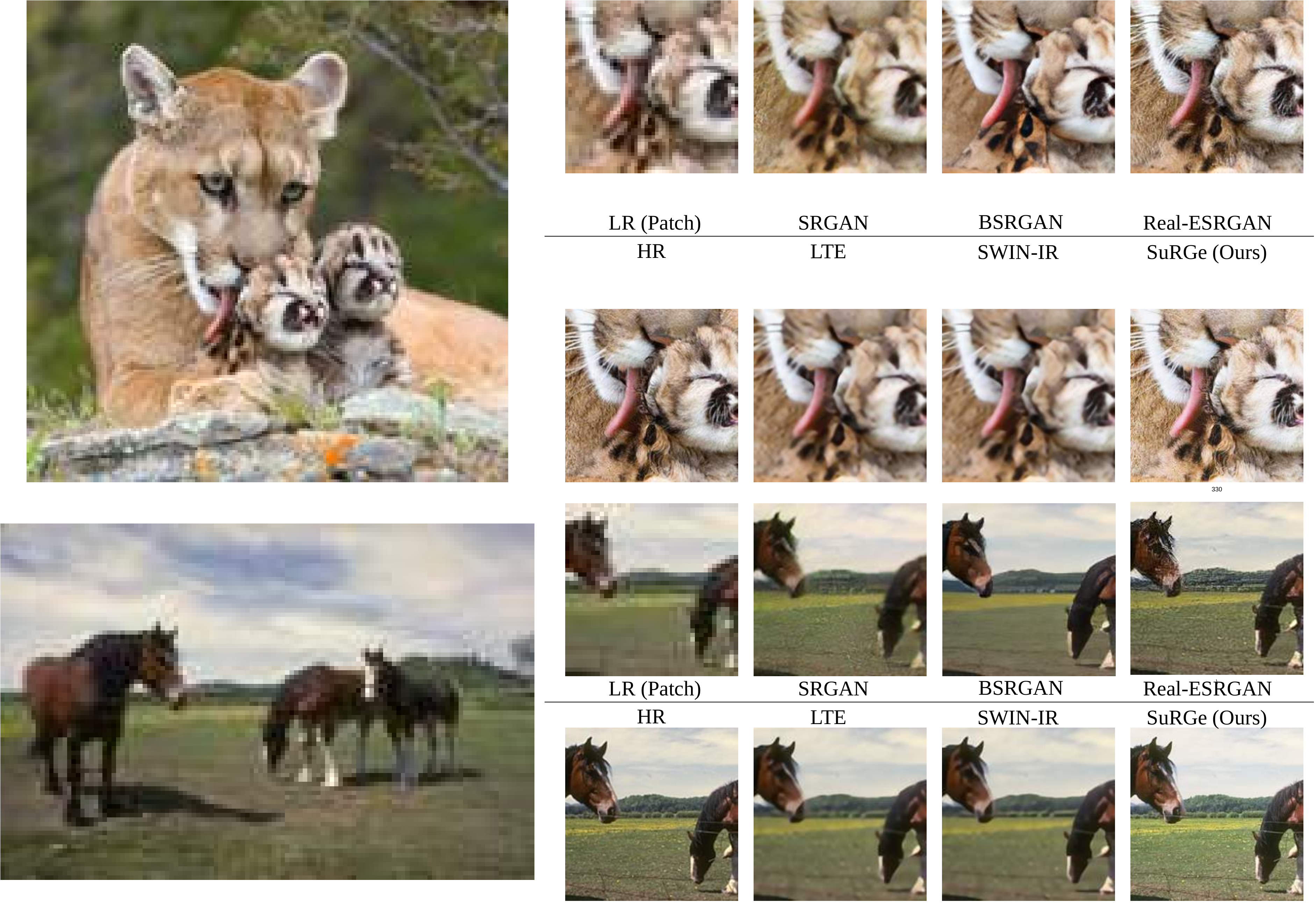}
    \caption{In continuation of Figure \ref{fig:addtional_qual}, we take two more test samples, namely Lioness (OST300) and Horses (BSD100), and compare the SR outputs of the same six methods. Here also, we see that SuRGe retains the finer details such as the fur of the lioness and her cubs, or the mane of the horses.}
    \label{fig:additonal_qual_extra}
\end{figure*}

Additionally, Figure \ref{fig:qual} shows a qualitative performance comparison of SuRGe with five notable contenders, namely SRGAN \cite{srgan_cvpr2017}, BSRGAN \cite{gu2019bsrgan}, Real-ESRGAN \cite{wang2021realesrgan}, LTE \cite{lee2022lte}, and SWIN-IR \cite{swiniriccvw2021}, on five test images, viz. Cars (PIRM), Statues (BSD100), Balloons (PIRM), Horses (BSD100), and Lioness (OST300). For a simultaneous quantitative comparison, we present the SSIM and PSNR values for the six algorithms on the five samples in Table \ref{tab:add_quant}. Figure \ref{fig:qual} demonstrates how SuRGe maintains a higher degree of details in the SR output, while the same is further quantitatively attested by the improved PSNR and SSIM in Table \ref{tab:add_quant}. Moreover, for the ease of visualization, a patch-based qualitative comparison is also presented in Figures \ref{fig:addtional_qual} and \ref{fig:additonal_qual_extra} that enables focused observation on a particular region of the SR output. 
\begin{table}[!ht]
    \centering
    \scriptsize
    \caption{Additional quantitative performance comparison of SuRGe with the five contenders and five test samples used in Figure \ref{fig:qual}. The best is boldfaced while the second-best is underlined.}
    \label{tab:add_quant}
    \vspace{5pt}
    \begin{tabular}{l|l|ccccc} \toprule
        Method & Metric & Cars & Statues & Balloons & Horses & Lioness \\ \midrule
         \multirow{2}{*}{SRGAN} & PSNR & 29.52 & 28.83 & 29.77 & 29.00 & 30.54 \\
         & SSIM & 0.57 & 0.46 & 0.65 & 0.59 & 0.71 \\ \midrule
         \multirow{2}{*}{BSRGAN} & PSNR & 30.29 & 29.29 & 30.38 & 31.05 & 30.67 \\
         & SSIM & 0.53 & 0.49 & 0.61 & 0.63 & 0.70 \\ \midrule 
         \multirow{2}{*}{Real-ESRGAN} & PSNR & 30.38 & 29.10 & 30.32 & 30.92 & 31.08 \\
         & SSIM & 0.58 & 0.48 & 0.65 & 0.62 & 0.70 \\ \midrule
         \multirow{2}{*}{LTE} & PSNR & 31.56 & 29.61 & 30.51 & 31.83 & 32.07 \\ 
         & SSIM & 0.71 & 0.58 & 0.79 & 0.70 & \underline{0.79} \\ \midrule
         \multirow{2}{*}{SWIN-IR} & PSNR & \underline{31.69} & \underline{29.65} & \underline{31.66} & \underline{31.93} & \underline{32.13} \\
         & SSIM & \underline{0.72} & \underline{0.59} & \underline{0.81} & \underline{0.72} & \underline{0.79} \\ \midrule 
         \multirow{2}{*}{SuRGe (ours)} & PSNR & \textbf{34.21} & \textbf{32.79} & \textbf{32.31} & \textbf{33.13} & \textbf{34.74} \\
         & SSIM & \textbf{0.82} & \textbf{0.79} & \textbf{0.86} & \textbf{0.86} & \textbf{0.91} \\ 
         \bottomrule 
    \end{tabular}  
\end{table}

\end{document}